\documentstyle[12pt,psfig,lscape]{l-aa}
\input{psfig}
\newcommand{\be}{\begin{equation}}
\newcommand{\ee}{\end{equation}}

\newcommand{\chivec}{{\mbox{\boldmath$\chi$}}}
\newcommand{\gammavec}{{\mbox{\boldmath$\gamma$}}}
\newcommand{\qvec}{{\bf{q}}}
\newcommand{\gvec}{{\bf{g}}}

\newcommand{\arcsecf}{{\hbox{$.\!\!^{\prime\prime}$}}}
\newcommand{\arcminf}{{\hbox{$.\!\!^{\prime}$}}}

\begin{document}
   \thesaurus{02         
              (12.03.4;
               12.07.1)}

\title{How accurately can we measure weak gravitational shear?}
\author{T. Erben$^{1}$, L. Van Waerbeke$^{2}$, E. Bertin$^{3,4}$, Y.
Mellier$^{3,4}$, P. Schneider$^{1,5}$}
\offprints{erben@mpa-garching.mpg.de}
\institute{
$^1$ Max Planck Institut f\"ur Astrophysik, Karl-Schwarzschild-Str. 1,
Postfach 1317, D-85741 Garching, Germany. \\
$^2$ Canadian Institut for Theoretical Astrophysics, 60 St
Georges Str., Toronto, M5S 3H8 Ontario, Canada.\\
$^3$ Institut d'Astrophysique de Paris. 98 bis, boulevard
Arago. 75014 Paris, France. \\
$^4$ Observatoire de Paris. DEMIRM. 61, avenue de
l'Observatoire.  75014 Paris, France.\\
$^5$ Universit\"at Bonn, Auf dem H\"ugel 71, D-53121 Bonn, Germany\\
}


\maketitle
   \markboth{How accurately can we measure weak gravitational shear?}{}

\begin{abstract}
With the recent detection of cosmic shear, the most challenging
effect of weak gravitational lensing has been observed. The main
difficulties for this detection were the need for
a large amount of high quality data and the control of systematics
during the gravitational shear measurement process, in particular
those coming from the Point Spread Function anisotropy.

In this paper we perform detailed simulations with the state-of-the-art
algorithm developed by Kaiser, Squires and Broadhurst (KSB) 
to measure gravitational shear. We show that
for realistic PSF profiles the KSB algorithm can recover any shear
amplitude in the range $0.012 < |\gammavec |<0.32$  with a relative,
systematic error of 
$10-15\%$.
We give quantitative limits on the PSF correction method
as a function of shear strength, object size, signal-to-noise and
PSF anisotropy amplitude, and we provide an automatic procedure to
get a reliable object
catalog for shear measurements out of the raw images.
\end{abstract}
\section{Introduction}
Weak gravitational lensing has become one of the most important
cosmological tools in the last decade. The mass distribution in the
universe can in principle be mapped from the measurement of the
coherent distortion caused by inhomogeneously distributed foreground
mass on the orientation of the faint background galaxies.  Lensing by
clusters of galaxies, strong enough to be routinely detected now,
already provides important constraints on the nature of our universe
[e.g. it probed massive high-redshift clusters (Luppino \& Kaiser
1997; LK97 henceforth; Clowe et al. 1998) and revealed the possible
existence of dark clumps (Erben et al. 2000, Umetsu et al. 2000)].
Since the first publications of ``cluster lensing'' (see e.g., Tyson,
Valdes \& Wenk 1990 and Kaiser \&
Squires 1993) much theoretical and observational progress on weak
lensing has been made, and interest progressively turned to very
weak shear measurements.  Galaxy-galaxy lensing is an example of such
a study which provides constraints on average dark matter halo
properties of galaxies, and cosmic shear is, ultimately, the direct 
measurement of
statistical properties of the large-scale matter distribution in our
universe (see, e.g., Schneider \& Rix 1997, Schneider et al. 1998b).
First galaxy-galaxy lensing experiments investigating dark halos of
field and cluster galaxies have been successfully made (Brainerd,
Blandford \& Smail 1995, Fischer et al. 1999, Natarayan et al. 1998). Very recently,
the cosmic shear was also detected (Schneider et al. 1998a, van Waerbeke et
al. 2000, Bacon et al. 2000a, Wittmann et al. 2000, Kaiser et
al. 2000). Because of the small amplitude of the lens effect, a
quantitative analysis of the signal requires high-quality data and
very precise data analysis for the detection of the signal.  With the
advent of new high-quality wide field CCD cameras substantial amounts of
useable data for these studies will soon become available [e.g., the
Descart project
(http://terapix.iap.fr/Descart/Descart\_english.html)].  

On the analysis side, the key issue is the determination of the
gravitational shear $\gammavec$ measured from galaxy shapes. Using
real data, we have to deal with observational and data reduction
defects. On CCD images, the information from very faint and small
objects whose shape we want to measure is often contained in only a
few image pixels and a clear relation of pixels to individual objects
is hard to achieve. Moreover, the measurements are rendered more
difficult by pixel noise and PSF effects that can mimic a possible
shear signal. Kaiser, Squires \& Broadhurst (1995, KSB henceforth)
have proposed a method for shear determination which is optimised for
the analysis of faint and small objects by utilising weighted
brightness moments of the light distribution. Their formalism (the
so-called IMCAT\footnote{see Nick Kaiser's home page,
http://www.ifa.hawaii.edu/$\sim$kaiser}) is to date one of the few
taking into account smearing and anisotropy effects from an
atmospheric PSF. Their formulae were derived for the weak-shear limit
and assuming that PSFs can be written as an isotropic PSF convolved
with a compact, anisotropic kernel.  In this paper we investigate the
accuracy and limitations of this method with simulations. We address
the following questions: (1) It was shown in earlier publications that
PSF profiles, for which the KSB breaks down, can be constructed very
easily (Kuijken 1999). So, is the proposed PSF correction procedure
valid for PSF profiles that we observe in ground-based observations?
(2) What influence does pixel noise have on the shear estimation? (3)
The KSB formulae are valid to first order in the shear ($\kappa\ll 1,
\;|\gammavec |\ll 1$). Is this expansion valid for typical weak
lensing applications with ground-based observations (from $|\gammavec
|=0.01$ for cosmic shear up to $|\gammavec |=0.2$ for cluster mass
reconstructions) or should we expand the formalism to higher order?
(4) Can we set up a fully automatic procedure, from reduced images to
an object catalog, for reliable shear measurements?

We generate
a large number of simulated images and analyse them exactly in the
same way as real data. We perform two kinds of simulations:
\begin{itemize}
\item Simulations where all objects involved (galaxies as well
as stars) have Gaussian profiles.
\item Simulations with the {\sl SkyMaker} program (Bertin 2000, in
preparation). This program generates galaxies modelled as exponential
disks with a central de Vaucouleurs type bulge of varying ratio. The
most important feature of SkyMaker is its ability to produce realistic
ground-based PSFs and to allow for the inclusion of several
telescope/detector defects (like astigmatism, coma, drift)
\end{itemize} 
The outline of the paper is as follows:
Sect. 2 summarises the KSB formalism and how we use it to estimate
gravitational shear. Sect. 3 gives a short overview over the
{\sl Stuff} program which generates the input galaxy and star catalogs
for the SkyMaker program. We describe our simulations with Gaussian
profiles in Sect. 4, our object detection and selection procedure
in Sect. 5. The SkyMaker simulations follow in Sect. 6 
and we finish with our conclusions in Sect. 7. The generation of
Point Spread Functions in the SkyMaker program is described in an
Appendix.
\section{Shear estimates}
\label{shearestsec}
KSB estimates the shear by considering the first order effects of a
gravitational shear and an instrumental Point Spread Function (PSF) on
the (complex) ellipticity of the galaxies $\chivec=\chi_1+{\rm i}\chi_2$,
which is defined as:
\be
  \chi_1=\frac{Q_{11}-Q_{22}}{Q_{11}+Q_{22}};
  \chi_2=\frac{2Q_{12}}{Q_{11}+Q_{22}},
\ee 
where the $Q_{ij}$ are weighted second brightness moments of the light
distribution $I$:
\be
  Q_{ij}=\int {\rm d}^2\theta\theta_i\theta_j I({\theta}) W(|\theta|),
\ee
where the object centre is at the origin of the coordinate system,
i.e.
\be
  \int {\rm d}^2\theta\theta I({\theta}) W(|\theta|)=0.
\ee
The total response of the galaxy ellipticity $\hat{\chivec}^0$, that is
the intrinsic ellipticity convolved with an isotropic function
[see Bartelmann \& Schneider (2000)], to a reduced gravitational shear
$\gvec=\gammavec/(1-\kappa)$ and the PSF is given by:
\be
  \label{correctionformula}
  \chivec-\hat{\chivec}^0=P^{\rm g}\gvec -P^{\rm sm}\qvec^*; \;\; 
  P^{\rm g}=P^{\rm sh}-P^{\rm sm}(P^{* \rm sm})^{-1}P^{* \rm sh},
\ee
where $\chivec$ is the observed ellipticity and the tensors 
$P^{\rm sh}$ and $P^{\rm sm}$ can directly be calculated
from the galaxy's light profile and the weight function $W$.
$P^{\rm sh}$ ({\sl Shear polarizability}) would be the response of 
the weighted galaxy ellipticity to a gravitational shear in the 
absence of PSF effects. $P^{\rm g}$ modifies this tensor by a factor
including the {\sl Smear polarizability} tensor $P^{\rm sm}$ 
to calibrate the shear estimate for the circular smearing by the PSF. 
This calibration also depends on the corresponding tensors $P^{* \rm sh}$ 
and $P^{* \rm sm}$ from stellar
objects containing the information of the PSF.
The stellar anisotropy kernel $\qvec^*$, needed for the correction of
the PSF anisotropy, can be estimated by noting
that $\hat{\chivec^0}^*=0$, $\gvec^*=0$ for stars so that
\be
  \qvec^* = (P^{* \rm sm})^{-1}\chivec^*.
\ee
A complete derivation of these formulae clarifying all the assumptions
in the formalism is published in Bartelmann \& Schneider (2000) but
see also KSB, LK97 and Hoekstra et al. (1998). 
To estimate $\gvec$ we first correct objects for the PSF anisotropy:
\be
  \label{anisotropycorrection}
  \chivec^{\rm aniso}=\chivec+P^{\rm sm}\qvec^*
\ee
and then consider averages over galaxy images in areas
where we assume constant reduced shear:
\be
  \langle\chivec^{\rm aniso}\rangle-\langle\hat{\chivec^0}\rangle= \langle
  P^{\rm g} \gvec\rangle= \langle P^{\rm g}\rangle\langle
  \gvec\rangle\rightarrow \langle \gvec\rangle=
  {\langle P^{\rm g}\rangle^{-1}\langle\chivec^{\rm aniso}\rangle};\;\;
  \label{shearestimategood}
\ee
where we used $\langle\hat{\chivec^0}\rangle=0$. For
simplicity eq. (\ref{correctionformula}) is often used in the form:
\be
  \gvec={(P^{\rm g})^{-1}}({\chivec^{\rm aniso}-\hat{\chivec^0}})
  \Rightarrow 
  \langle\gvec\rangle = 
  \left\langle{(P^{\rm g})^{-1}}{\chivec^{\rm aniso}}\right\rangle
  \label{shearestimatebad}
\ee
where we have to assume
$\langle{(P^{\rm g})^{-1}}{\hat{\chivec^0}}\rangle=0$.
We checked that with our procedure described below this assumption holds and
the estimators
from eq. (\ref{shearestimategood}) and eq. (\ref{shearestimatebad})
are in very good agreement, although in practice the latter is easier to handle.
\subsection{Practical application of the KSB formulae}
\label{applicationref}
Although the general KSB procedure seems uniquely determined, there are
several possibilities for its practical implementation.
This has been the subject of discussions in recent works (Hoekstra et
al. 1998, Kuijken 1999, Kaiser 1999) on how the KSB formulae should be
applied optimally.
One issue is the window function $W(|\theta |)$ with which galaxies
are weighted. KSB have chosen a Gaussian with a scale $\sigma$
proportional to the size of the object under consideration. Hoekstra
et al. (1998) noted first that quantities from stellar objects should be
calculated with the same scale as the object to be corrected (this is
the most natural choice although according to the derivation
of the KSB formulation the result should not depend on it).
Another difficulty arises as the quantities $\qvec^*$ and
$(P^{* \rm sm})^{-1}P^{* \rm sh}$ have to be known at the positions of
the galaxies to
be corrected. Since these quantities are known only at star locations, we need
to estimate what the PSF would be on the position of each detected galaxy.
On top of all, all measured quantities are affected by noise in an unknown way.
Two techniques have been proposed to suppress this noise in the shear
estimate. First, the noise in the scaling factor $P^{\rm g}$
can be minimised by a fit or by calculating means or medians in
bins in the parameter space $\pi_{P^{\rm g}}$=(object size,
magnitude) which has often been used in the literature.
Second, we can easily introduce a weighting scheme in the
estimators (\ref{shearestimategood}) and (\ref{shearestimatebad}).

\begin{figure*}[t]
  \centerline{\psfig{figure=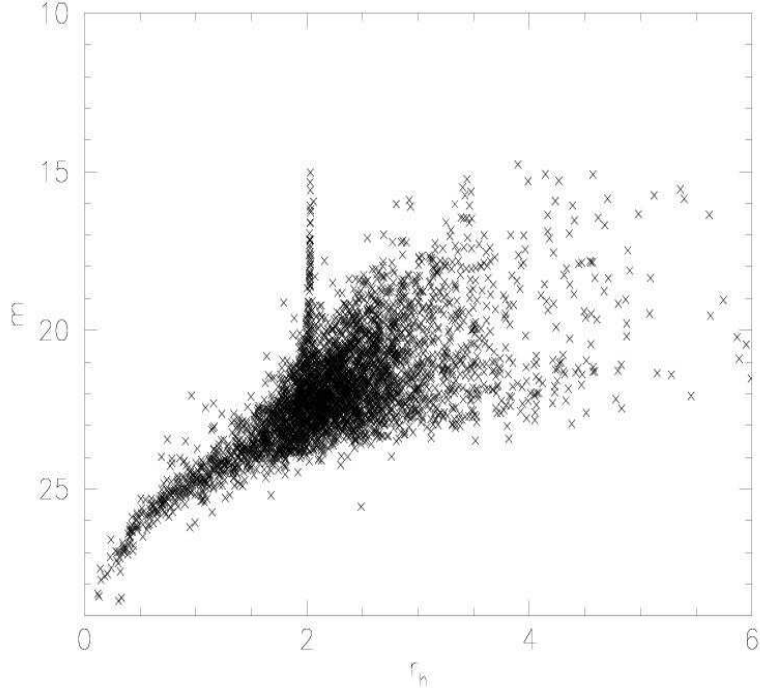,height=0.4\textheight,width=0.6\textwidth}}
  \caption{Plotted is the half light radius vs. the magnitude (in
  arbitrary units) from all the detections in one of our SkyMaker
  simulations in Sect. \ref{skymakersimulsec}. Bright,
  unsaturated stars are clearly localised in a small region around
  $r_h\approx 2$ and in $15<{\rm m}<20$. Larger objects are considered
  as
  galaxies, while the smaller, very faint objects are probably noise
  detections. For objects around the stellar locus with faint
  magnitudes, a clear classification between stars and galaxies cannot
  be done and we typically exclude this region from our analysis.}
  \label{rhmagfig}
\end{figure*}

Our procedure, which we found to be optimal in terms of shear recovery
accuracy, can be summarised as follows:
To apply (\ref{correctionformula}), for every galaxy the quantities
$\qvec^*$ and $(P^{* \rm sm})^{-1}P^{* \rm sh}$ have to be known everywhere on the
image.  Usually the PSF anisotropy varies smoothly over the field so
that the $q^*_i$ can be well represented by a low-order polynomial fit from
the light profiles of bright, unsaturated foreground stars. A second-order polynomial
is enough, and higher order polynomials do not improve the PSF correction. We
select these stars in a $r_{\rm h}$ vs. m diagram (see
Fig. \ref{rhmagfig}),
and then measure $\qvec^*$ with a filter of the size of the stars, but
$(P^{* \rm sm})^{-1}P^{* \rm sh}$ in a range of filter scales spanning the sizes of
our galaxies. We then perform the anisotropy correction
(\ref{anisotropycorrection}) for each individual galaxy.
For $P^{\rm g}$ we finally use the raw, unsmoothed values as our analysis
shows that fitting this quantity does not improve the final results
(see Sect. \ref{numericsection}).

Corrected galaxy ellipticities are subject to high noise, which can
eventually produce unphysical ellipticities much larger than unity. 
It is therefore necessary
to weight each galaxy according to the accuracy of its ellipticity
measurement/correction.
For our weighting scheme we assume the ellipticity distribution
of the lensed galaxies is to first order equivalent to the intrinsic
ellipticity distribution everywhere on our data field. 
We also assume that the intrinsic ellipticity distribution does not
vary over the field. So, high ellipticities are assumed to
originate from noise and these galaxies should be assigned a low
weight. These assumptions
are valid for studies of {\sl empty fields} but not when
investigating cluster lenses where high ellipticities are lensing features
and not necessarily caused by noise. In this case a weighting scheme calculating
weights out of the pixel noise properties may be adopted. 
See the Appendix of Hoekstra et al. (1999) for an example.

We introduce a simple $U=1/\sigma_{\rm g}^2$
weighting, where $\sigma_{\rm g}^2$ is the variance of the shear estimators
from individual objects. These weights are estimated in a
parameter space $\pi_U$ supposed to trace the noise properties of
objects (like object size and S/N for example).
Unfortunately our objects are typically
very clustered in $\pi_U$ so that calculating
averages in cells defined by a regular grid is not optimal and an
adaptive grid should be more appropriate. Therefore for each galaxy we consider  
its $N$ nearest neighbours 
in our parameter space (typically $N\simeq 20$) and assign the inverse of
$\sigma_{\rm g}^2=\frac 1N\sum |\gvec|^2$ from these neighbours 
as weight to that galaxy. The distance $d$ from a galaxy $k$ to its
neighbour $l$ in a parameter space $\pi$ consisting of $i=1..M$ elements 
is hereby defined by
\be
  d=\sqrt{\sum_{i=1}^M (\pi_i^k-\pi_i^l)^2}. 
\ee
Obviously, a galaxy
does not have an unique set of closest neighbours in an
arbitrary space.
For instance, object sizes would simply scale by a factor of 2 if
data are rebinned to half the original resolution while quantities like m stay
unchanged. Possible consequences of this are not investigated here.
Hereafter we denote averages $\langle x\rangle$ and uncertainties of 
means $\sigma_{\langle x\rangle}$ from a weighted quantity $x$ by:
\begin{eqnarray}
  \label{weightmean}
  \langle x\rangle &=& \frac{\sum_k x_k U_k}{\sum_k U_k}; \\
  \sigma_{\langle x\rangle}^2 &=& \frac{1}{\sum_k U_k}. \nonumber
\end{eqnarray}
As noted above, our weighting scheme is well justified only 
if the lensed ellipticity
distribution is to first order equivalent to the unlensed one.
We found in our simulations that we can also estimate shear
that is not small with respect to the intrinsic ellipticity
distribution with the above formulae but not the corresponding
errors. The reason
is that only relative weights enter the shear estimate while
the errors $\sigma_{\langle x\rangle}^2$ are determined by the 
absolute values of the weights.
In the strong shear case, that is, if the intrinsic ellipticity distribution
is substantially changed by the input shear, the absolute values
of the weights directly depend on the input shear.
Hence we first use eq. (\ref{weightmean}) to estimate the reduced shear
according to (\ref{shearestimatebad}):
\be
  g_{\alpha}=\langle (P^{\rm g})^{-1}_{\alpha\beta}\chi^{\rm aniso}_{\beta}\rangle,
  \label{tensorpgamma}
\ee
and then estimate the errors by bootstrapping.
In parallel to using the full tensor expression for $P^{\rm g}$ we also calculate
the often used scalar correction, where $P^{\rm g}$ is estimated by
$P^{\rm g}_s$:
\be
  P^{\rm g}_s=0.5\; {\rm tr} [P^{\rm g}]\; ;
  \left\langle \gvec\right\rangle= \left\langle\frac{\chivec^{\rm aniso}}{
  {P^{\rm g}_s}}\right\rangle,
  \label{scalarpgamma} 
\ee 
(Hudson et al. 1998, Hoekstra et al. 1998). The weights for the tensor
and scalar case are generally different.
\section{Catalog generation}
\label{stuffsec}
For our simulations we used the Stuff program\footnote{Freely available
at: {\tt ftp://ftp.iap.fr/pub/from\_users/bertin/stuff/}} to generate our
initial object catalogs. Details of this program and the
SkyMaker tool producing images out of these catalogs will be
published elsewhere (Bertin 2000, in preparation); here, only a
very short summary of those parts relevant for this work are given:

Given
the sky dimensions of the intended simulations, the program distributes
galaxies in redshift space that is subdivided into bins. For every bin,
representing a volume element according to the specified cosmology
(we used $H_0=65$km/(s Mpc), $\Omega_{\rm m}=0.3$ and $\Omega_{\lambda}=0.7$
in all our simulations), the
number of galaxies for the Hubble types E, S0, Sab, Sbc, Scd and
Sdm/Irr is determined from a Poisson distribution assuming a
non-evolving 
Schechter luminosity function (Schechter 1976).
The
different galaxy types are simulated by linearly adding exponential
$\mu_{\rm d}(r)$ (disk component) and de Vaucouleur profiles $\mu_{\rm b}(r)$
(bulge components) in different ratios.  
\begin{eqnarray}
\mu_{\rm b}(r)&=& M_{\rm b}+8.3268\left(\frac r{r_{\rm
b}}\right)^{1/4}+5\log r_{\rm b}+ 16.6337 \\
\mu_{\rm d}(r)&=& M_{\rm d}+1.8222\left(\frac r{r_{\rm d}}\right)+
5\log r_{\rm d}+0.8710,
\end{eqnarray}
where $\mu_{\rm b}$, $\mu_{\rm d}$ are the surface brightnesses in
mag/pc$^2$, and $M_{\rm b}$, $M_{\rm d}$ are the absolute magnitudes of the bulge and
the disk components, respectively. The distributions of scale radii
$r_{\rm b}$ and $r_{\rm d}$ are fixed by
an empirical relation (Binggeli et al. 1994) and a semi-analytical
model (de Jong \& Lacey 1999) relating these quantities to the
absolute magnitudes. The galaxies are assigned a random disk
inclination angle and a position angle which define the intrinsic
ellipticities of our objects. The output of the program is a catalog
of galaxy positions, apparent magnitudes, semi-minor and major axes,
and position angles for disks and bulges. These catalogs are then
sheared (disks and bulges are sheared separately!), processed with
SkyMaker and run through our KSB procedure. Before describing
these SkyMaker simulations we present results with Gaussian object
profiles in the next section.
\section{Semi-analytical calculations with Gaussian profiles}
\label{numericsection}
Although Gaussian profiles do not fit real galaxies and stars very
well, they allow a quick investigation of the possible biases
connected with the KSB procedure that arise from constraints in real
data. Moreover most of the quantities defined in KSB can be
analytically calculated with Gaussian profiles, which allows one to check
that the numerical simulations are done properly and to explore the
following effects:
\begin{itemize}
\item CCD images are pixelised and all integrals involved in the KSB
procedure have to be estimated by discrete sums.
\item Our data are affected by sky noise. The expressions from
Sect. \ref{shearestsec} are non-linear in the object profile $I$, so
noise in this quantity may introduce systematic biases. Sky noise also
affects the calculation of the objects centroid position from which
all shear quantities are calculated. So we have to investigate how the
noise in object positions influences the final shear estimates.
\item In practice we only have the observed light distribution for our
analysis. When applying the KSB formalism we make the assumption that
the influence of the anisotropic component of the PSF and shearing on
the profiles are ``small'', although in principle intrinsic profiles
should be used.
\end{itemize}
Furthermore we investigate different possible parameter spaces
$\pi_{P^{\rm g}}$ and $\pi_U$ in view of a possible noise reduction in
$P^{\rm g}$ and our weighting scheme.  Our Gaussian galaxy profiles
were convolved with Gaussian PSFs given by:
\begin{equation}
I = A \exp(-\frac{x'^2}{a^2}-\frac{y'^2}{b^2}),
\end{equation}
where $x' = x\cos (\theta )-y\sin (\theta ), y'= x\sin
(\theta)+y\cos(\theta )$ for an object with semi-major (semi-minor)
axes $a$ and $b$ and a position angle $\theta$ with respect to the
abscissa. The dimensions of the pixels and the PSF profiles were
adapted to a CCD with 0\arcsecf 2 resolution and a seeing (FWHM) of
0\arcsecf 7 (This means $a_{*}=b_{*}=2.012$ pixel units for an
isotropic PSF).  We chose $A_{*} = 10000$ to get high S/N measurements
for all stellar quantities. The intrinsic scales for the galaxies $a$
and $b$ were taken from the output of the Stuff program. The
distributions of these scale radii are very peaked at around
$0.2\arcsec$ and have a long tail towards larger radii.
The formal mean intrinsic size (defined as $\sqrt{ab}$) of the
galaxies is $0.22\arcsec\pm 0.13\arcsec$.
The apparent magnitudes from Stuff were used to fix
the amplitude $A$. The centres of the objects were chosen randomly
within a pixel.

We then performed the KSB measurements on these
profiles for a noise-free, and for a Gaussian noise model with
$\sigma_{\rm sky}=13.0$ (this adapts these calculations to the
characteristics of the simulations in
Sect. \ref{skysimulsec}). Hereby we added the noise only to the
galaxies but not to the stars. So we can investigate the best possible
noise-free results only having the pixelisation as a source of error,
and the consequences of sky noise on them. In the following analysis
we used all the galaxies in the initial Stuff catalog regardless of
whether the objects would be found by some detection algorithm in a
real image or not.  For the weight function $W(|\theta |)$ we chose
here a Gaussian with scale $r_{\rm g}=1.107 \frac{\sqrt{2}a_{\rm
s}b_{\rm s}} {\sqrt{a_{\rm s}^2+b_{\rm s}^2}}$, where $a_{\rm s}$ and
$b_{\rm s}$ are the scale radii of our objects after smoothing with
the PSF. This corresponds to the $r_{\rm g}$ radius from KSB (see
Sect. \ref{selectionsec}) for a circular Gaussian $a_{\rm s}=b_{\rm
s}$.  We have checked that the results that we present in this section
do not depend significantly on this choice. Before calculating any
lensing quantities we estimated the object centre by iteratively
solving the equation:
\begin{equation}
\label{goodposeq}
\theta_0=\frac{\int\theta I(\theta)W(|\theta-\theta_0 |){\rm d}^2\theta}
             {\int I(\theta)W(|\theta-\theta_0 |){\rm d}^2\theta},
\end{equation}
where we used the pixel centre of the true position as starting point.
In addition, for every object we have defined a signal-to-noise
ratio by:
\begin{equation}
\label{SNratio}
\frac {\rm S}{\rm N} = \frac{\int I(\theta)W(|\theta |){\rm
d}^2\theta}{\sigma_{\rm sky}\sqrt{\int W^2(|\theta |){\rm d}^2\theta}}, 
\end{equation}
as we wanted a S/N ratio that is based on the same filter function 
as the one that the measurements are
done with (in the end it turned out that there is a linear relation between
our S/N estimator and the $\nu$ estimator from the KSB hierarchical
peak finder; see Sect. \ref{selectionsec}).
For the sample with S/N$ > 2$ the rms of $\Delta x$ and $\Delta y$
in the determination of the object centre is better than $1/20$ pixel.
For the low S/N$ < 2$ objects the rms is about 1 pixel in both
components.
All other quantities were then calculated with respect to
the estimated centre, where the integration was done up to $|\Delta
x|\leq 3 r_{\rm g};\; |\Delta y|\leq 3 r_{\rm g}$. 
We generated catalogs for the seven shear combinations listed in
Table \ref{summarytable} that cover the range from $|\gvec |=0.012$ to
$|\gvec |=0.32$.
\begin{table}
\begin{center}
\begin{tabular}{|c|c|c|c|c|c|c|c|}
\hline
$g_1$ & 0.01  & 0.03 & 0.15 & 0.25 & 0.08  & 0.2  & 0.14 \\ 
\hline
$g_2$ & 0.007 & 0.05 & 0.1  & 0.2 & 0.27 & 0.14 & 0.2 \\
\hline
\end{tabular}
\end{center}
\caption{\label{summarytable} The shear combinations we investigated
in this section with Gaussian profiles for the galaxies.}
\end{table}
For each combination we generated a final catalog with about 75000
objects having $15.0 < {\rm m} < 28.0$. 
If not stated otherwise our galaxies have the intrinsic
ellipticity generated by the Stuff program.
\subsection{Estimation of $P^{\rm g}$}
\begin{figure*}[ht]
  \psfig{figure=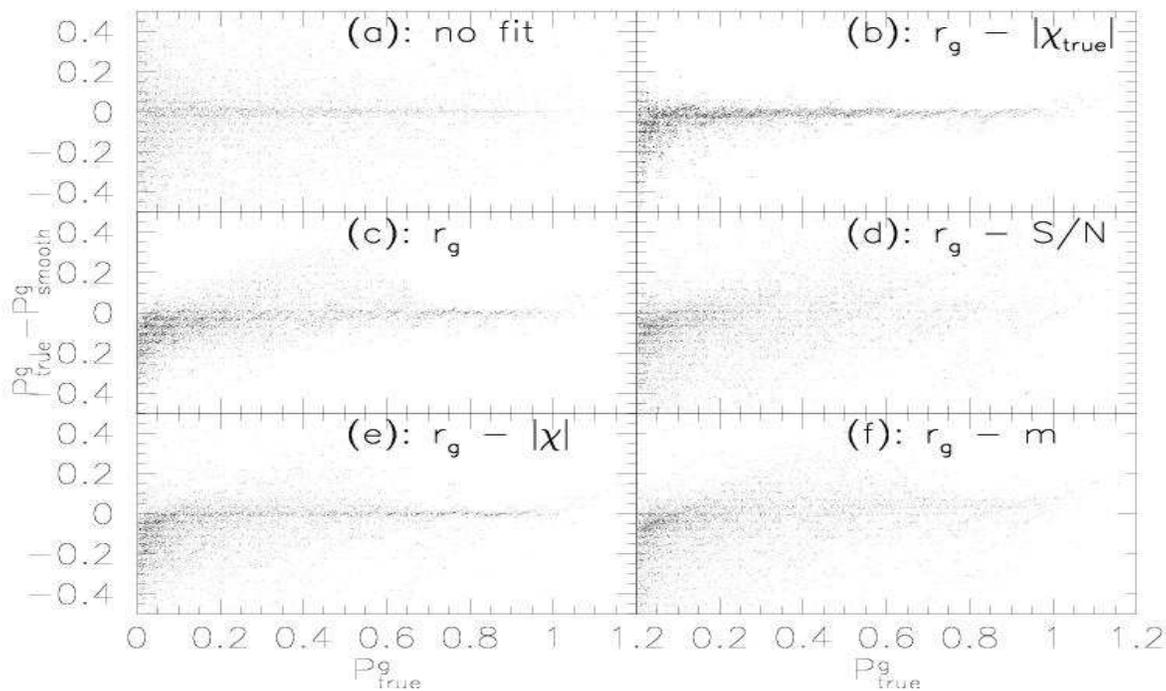,height=0.4\textheight,width=0.9\textwidth}
  \caption{The difference between the true and smoothed $P^{\rm g}$
           as function of $P^{\rm g}_{\rm true}$ with $P^{\rm g}_{\rm smooth}$
           calculated in different parameter spaces. In panel (a)
           with no smoothing the errors of $P^{\rm g}$ are randomly
           distributed around the true value. In 
           panel (c) (smoothing as function of $r_{\rm g}$) we see a
           systematic tail for intermediate values of
           $P^{\rm g}_{\rm true}$ indicating the dependence of
           $P^{\rm g}$ on at least one second parameter. Otherwise
           this smoothing reduces the noise very well. Panels (d) and (f)
           show that $P^{\rm g}$ does not
           depend on S/N or m. From panel (b) we see
           that $P^{\rm g}$ depends on the raw ellipticity $|\chi|$ as
           well as on $r_{\rm g}$.
           Unfortunately our measured, noisy
           ellipticities lead to significant noise in the smoothing
           [panel (e)].
 }
  \label{Pgammadifffig}
\end{figure*}

As we have calculated all quantities with and without sky noise
we can investigate whether the smoothed $P^{\rm g}$ in some
parameter space $\pi_{P^{\rm g}}$ can lower the noise in this quantity, and
whether this improves the final shear estimation.
For this we took the $g_1=0.25; g_2=0.2$ realisation and the 
scalar $P^{\rm g}$ estimator from eq. (\ref{scalarpgamma}).
We smoothed $P^{\rm g}$ with the next neighbour approach described
in Sect. \ref{applicationref} in several parameter
spaces to obtain $P^{\rm g}_{\rm smooth}$ and compared these values 
with the noise free $P^{\rm g}_{\rm true}$. Here we finally used the
median to estimate $P^{\rm g}_{\rm smooth}$ as this turned out to give
better results than the mean. The result is shown in Fig.
\ref{Pgammadifffig}.  We see that $P^{\rm g}$ is mainly a function of
the object size $r_{\rm g}$ and the modulus of the raw, noise-free but
unobservable ellipticity $\chi_{\rm true}$.  In practice, we only observe
noisy values for $\chi$ and a smoothing as a function of $r_{\rm g}$ alone
seems as good as smoothing in $r_{\rm g}$ and $|\chi|$.  Including quantities
like m or S/N is clearly worse. 

\begin{figure*}[ht]
  \centerline{\psfig{figure=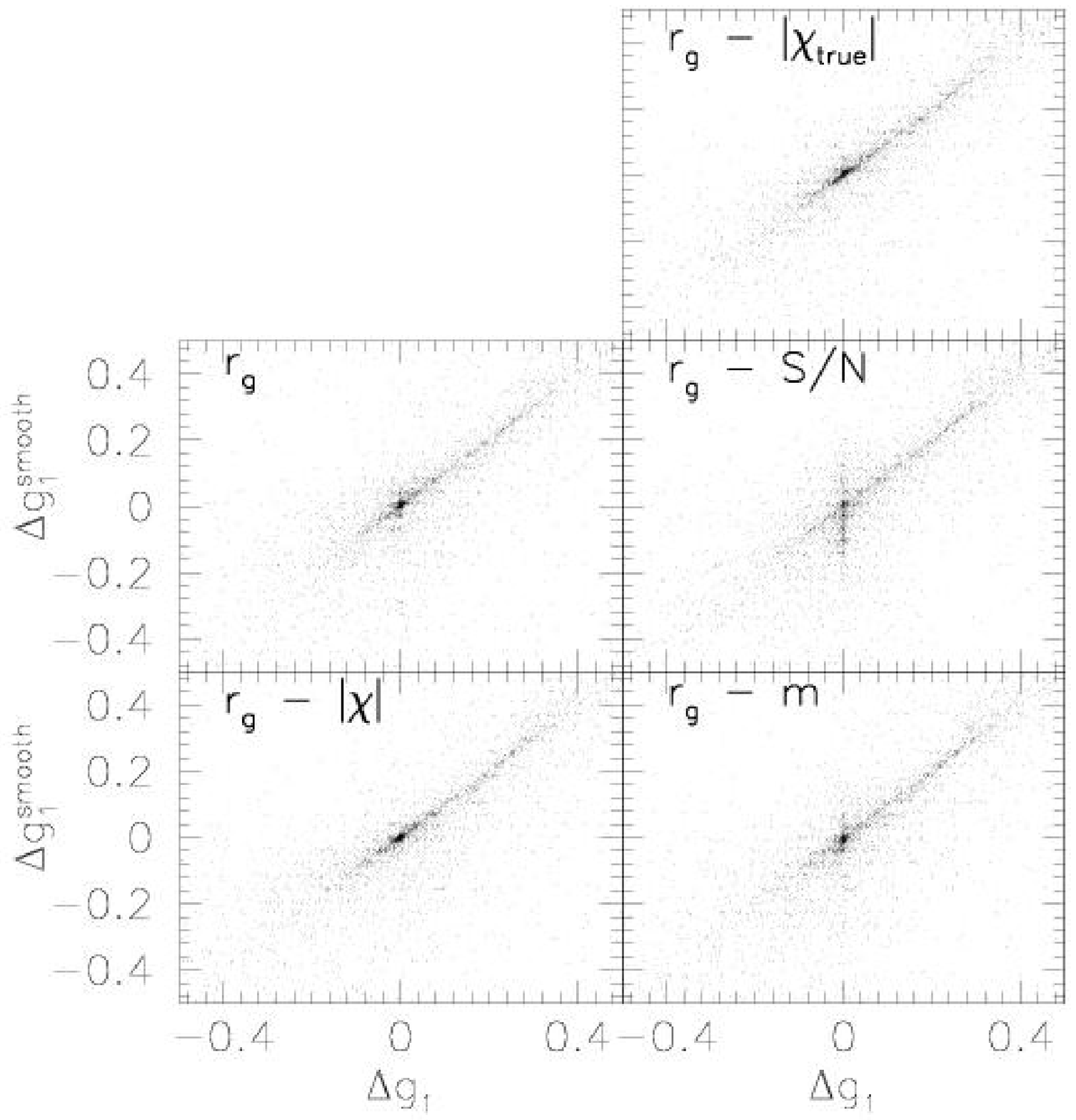,height=0.4\textheight}}
  \caption{The errors $\Delta g_1:=g_1^{\rm true}-g_1$ vs.
           $\Delta g_1^{\rm smooth}:=g_1^{\rm true}-g_1^{\rm smooth}$ for the various
           smoothings of $P^{\rm g}$. Even for the best smoothing
           of this quantity with $r_{\rm g}$ and the true raw ellipticity
           $\chi_{\rm true}$ we have $\Delta g\approx \Delta g^{\rm smooth}$
           indicating that smoothing $P^{\rm g}$ does not
           improve the shear estimation. The parameter spaces $r_{\rm g}$
           and $r_{\rm g} - |\chi|$ give similar results while 
           the $r_{\rm g}-$S/N smoothing especially shows a spread of 
           $\Delta g_1^{\rm smooth}$ where $\Delta g_1=0$. The result
           implies that smoothing $P^{\rm g}$ does not improve the final shear
           estimate. We checked
           that this general result also applies for objects with an
           intrinsic, unphysical negative value for $P^{\rm g}$ (see text). }
  \label{e1isodifffig}
\end{figure*}

In order to see whether these
smoothings improve the final shear estimates, we first calculated the
estimators $g = \frac{{\chi}} {P^{\rm g}}$ and
$g^{\rm smooth} = \frac{{\chi}}{P^{\rm g}_{\rm smooth}}$ for the various
smoothings. Fig. \ref{e1isodifffig} shows that even for the best
$r_{\rm g} - |\chi_{\rm true}|$ we find $\Delta g\approx\Delta g^{\rm
smooth}$. This shows that smoothing does not improve the shear
estimates on the whole but we still have to investigate whether it
helps for objects having an intrinsically unphysical negative $P^{\rm
g}$. For this we compared $\langle \Delta g\rangle$ and 
$\langle \Delta g^{\rm smooth}\rangle$ together with the corresponding
uncertainties for the subsamples with $P^{\rm g}>0$ and $P^{\rm g}<0$.
To exclude extreme outliers from this calculation we include only objects with S/N$>2.0$ and
a modulus smaller than unity for the corresponding shear estimate.
We obtain $\langle \Delta g^{\rm smooth}\rangle=0.028\pm 0.19$,
$\langle \Delta g\rangle=0.028\pm0.18$ for the $P^{\rm g}>0$
sample, and $\langle \Delta g^{\rm smooth}\rangle=0.16\pm 0.38$,
$\langle \Delta g\rangle=0.26\pm 0.42$ for $P^{\rm g}<0$
objects. This confirms that smoothing does not at all change the
estimate for $P^{\rm g}>0$. It also shows that we cannot recover the
shear signal with $P^{\rm g}<0$ objects, whether we use a smoothed
value for this quantity or not. 
\subsection{Weighting scheme}
\begin{figure*}[ht]
  \psfig{figure=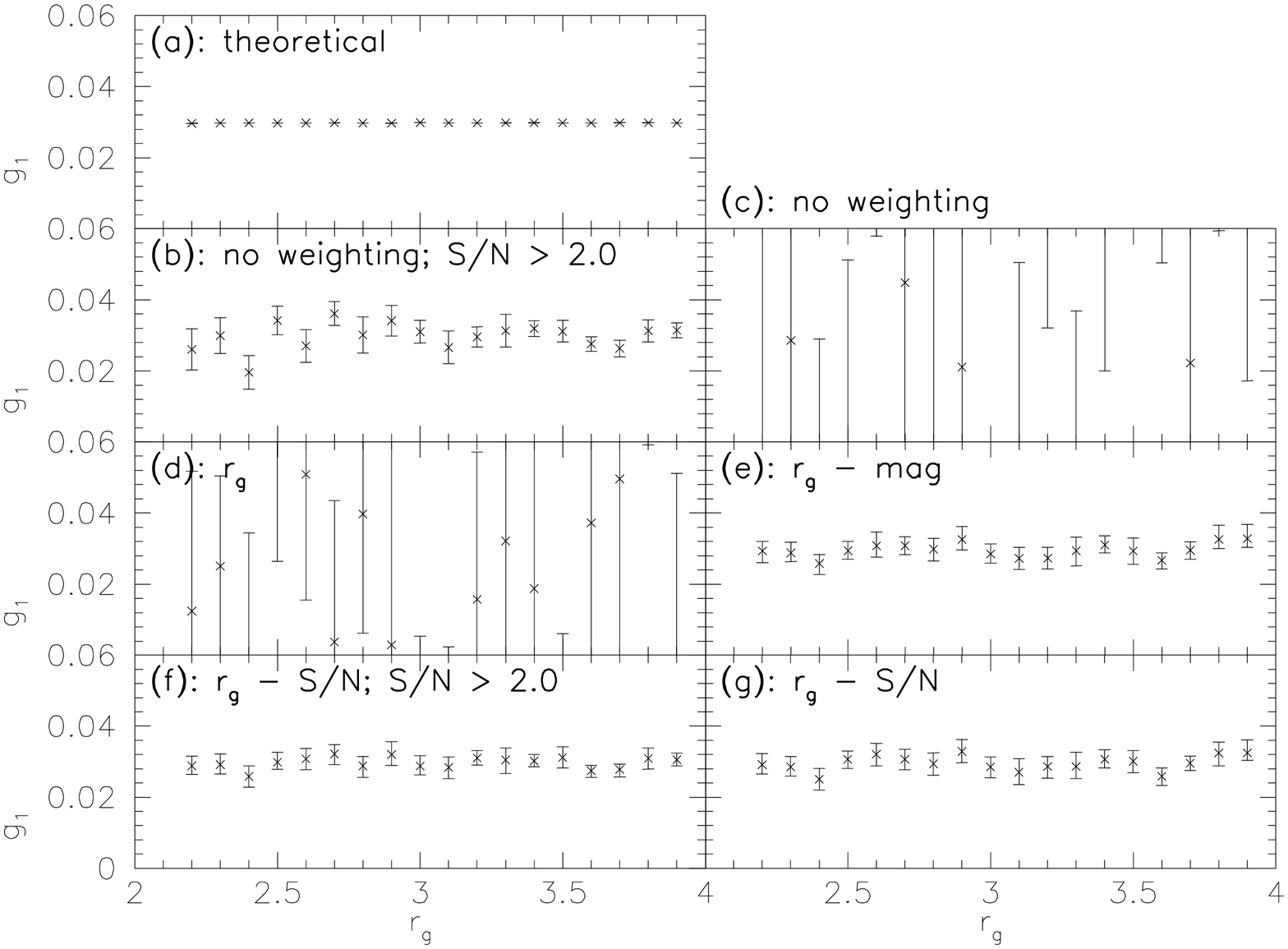,height=0.4\textheight,width=0.9\textwidth}
  \caption{The measured reduced shear $g_1$ as a function of the
  object size for various parameter spaces used in our weighting
  scheme. Panel (a) shows the result for noise free objects.
  Panels (b) and (c) show results if no weighting at all is applied,
  (d) and (e) for $r_{\rm g}$ and $r_{\rm g}-$m smoothings and finally
  (f) and (g) for the parameters $r_{\rm g}g-$S/N. 
  The error bars are from 200 bootstrap realisations of the galaxy
  samples (see text). We note that the
  weighting scheme improves the measurement significantly to the
  cases where no weighting scheme is applied. We also conclude that
  there is no need to exclude low S/N objects from the measurement if
  weighting is applied.
}
\label{weightbestfig}
\end{figure*}

Similarly, we now search for the
best possible parameter space to calculate $\sigma_{\rm g}^2$,
and thus the weighting of the galaxy ellipticities. 
For this investigation we used a catalog with intrinsically
round objects to isolate pixel noise errors from those of the
intrinsic ellipticity distribution. The input shear is $g_1=0.03$,
$g_2=0.0$.
Fig. \ref{weightbestfig} shows that using $\pi_U=(r_{\rm g},{\rm m})$
and $\pi_U=(r_{\rm g}, {\rm S/N})$ as parameter spaces for the
$\sigma_{\rm g}^2$ calculation give the best and very
comparable results in the final shear estimation. The errorbars in
this and all following calculations are estimated by creating
200 bootstrap realisations of our catalog in the following way: 
We randomly take $N$ objects out of our catalog containing $N$ objects
allowing for multiple entries. From this sample we recalculate a shear estimate
and repeat the exercise $N_{\rm boot}=200$ times. From these shear
estimates $g_i^{\rm boot}; i\in[1..N_{\rm boot}]$ the errorbar for
the original measurement is estimated by 
$\sigma_{\langle x\rangle}^2=\frac 1{N_{\rm boot}}\sum_i (g_i^{\rm boot}-\langle g_i^{\rm boot}\rangle)^2$
(we checked that $\langle g_i^{\rm boot}\rangle$ is in excellent agreement
with our original shear measurement). 
We note from Fig. \ref{weightbestfig} that introducing our weighting
scheme improves the shear
estimation significantly in comparison to no weighting.
Therefore, weights from the $\pi_U=(r_{\rm g},{\rm S/N})$ smoothing are now used throughout.
\subsection{Accuracy of shear estimates}
\begin{figure*}[ht]
  \psfig{figure=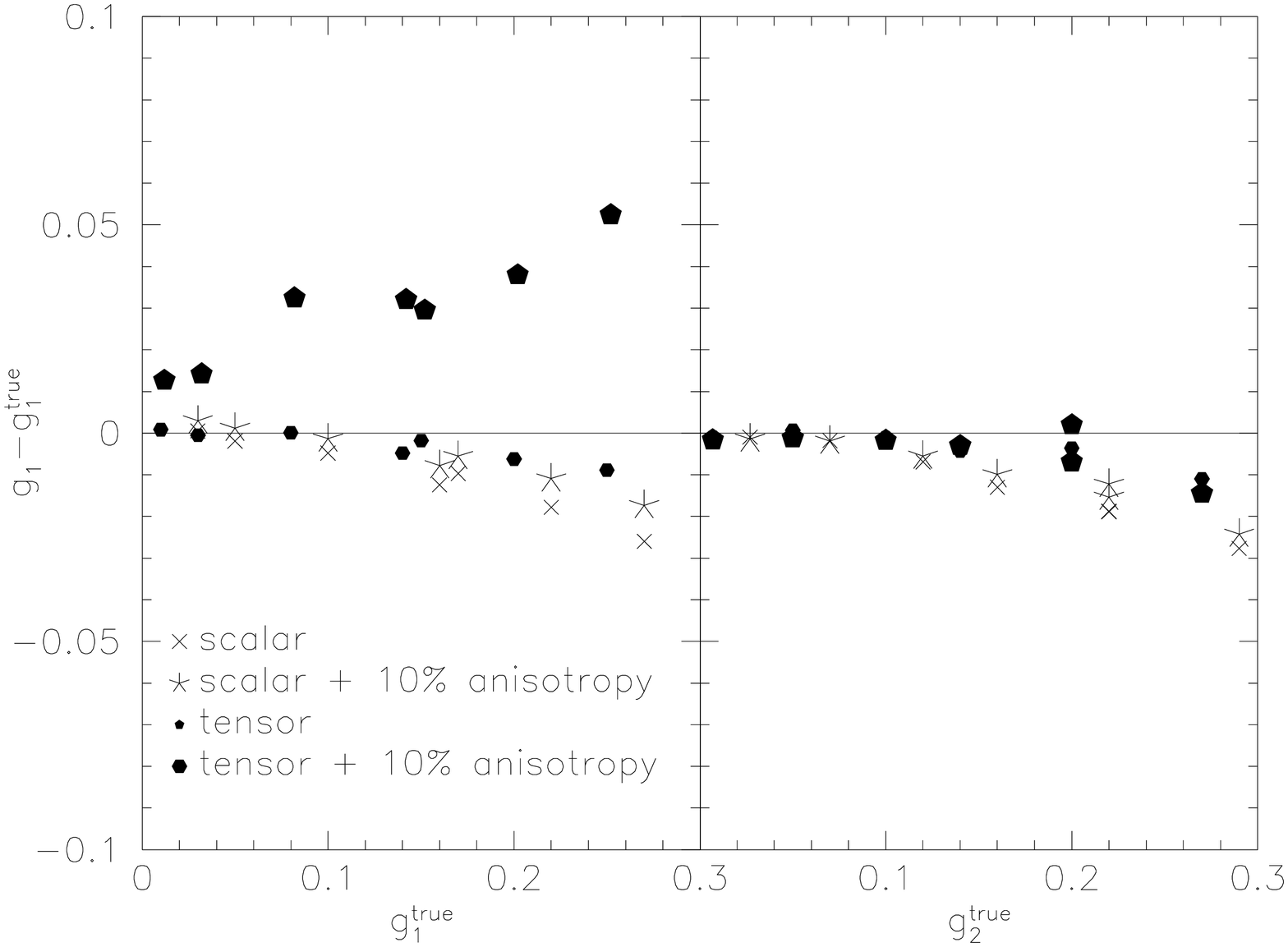,height=0.4\textheight,width=0.9\textwidth}
  \caption{The errors on shear recovery for noise-free
  objects. The crosses and filled hexagons show results for the scalar
  and tensor estimates with no PSF anisotropy present, the stars and
  filled pentagons for calculations with a $10\%$ PSF anisotropy in
  the $g_1$ direction. No object selection was done and our weighting
  scheme was used. The points were obtained by averaging over
  all $\approx 75000$ galaxies that we have for every shear
  realisation in our catalog. The error bars are less than 0.1 percent
  and so smaller than the symbols. For clarity
  of the plot the errorbars have been omitted and the symbols for the
  scalar correction have been slightly shifted to the right.}  
  \label{truevalues}
\end{figure*}

\begin{figure*}[ht]
  \psfig{figure=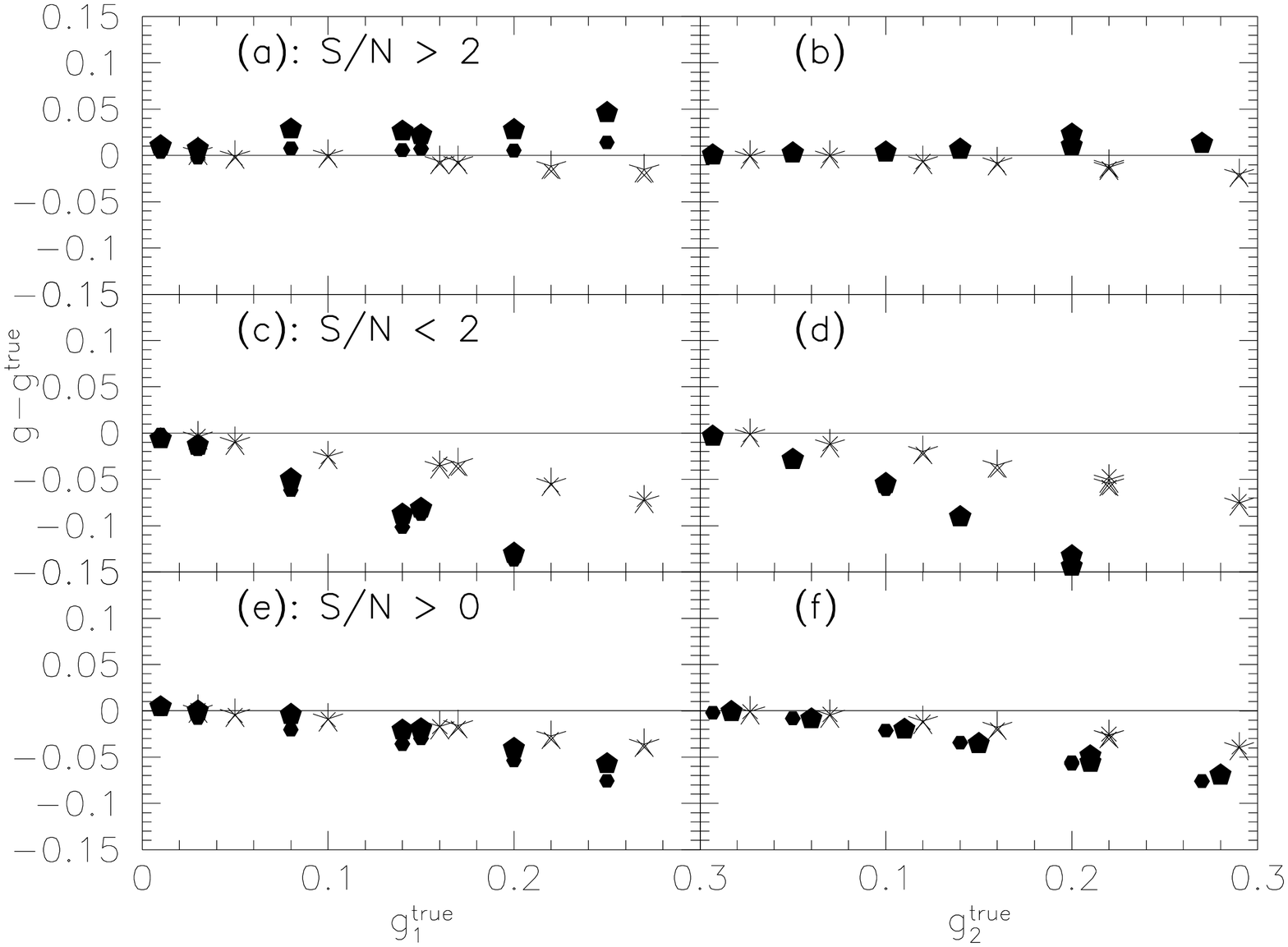,height=0.4\textheight,width=0.9\textwidth}
  \caption{The same as Fig. \ref{truevalues} for noisy objects.
  The two upper most panels show results for S/N$>2.0$, the middle
  panels for S/N$<2.0$ and the lower panels for the whole sample.
  For high S/N objects, the tensor estimation is more accurate but
  in general the scalar calculation is much more stable towards low
  signal objects. Symbols for the scalar correction are slightly
  shifted to the right for clarity of the plot. For more details see the text.}
  \label{measuredvalues}
\end{figure*}

\begin{figure*}[ht]
  \psfig{figure=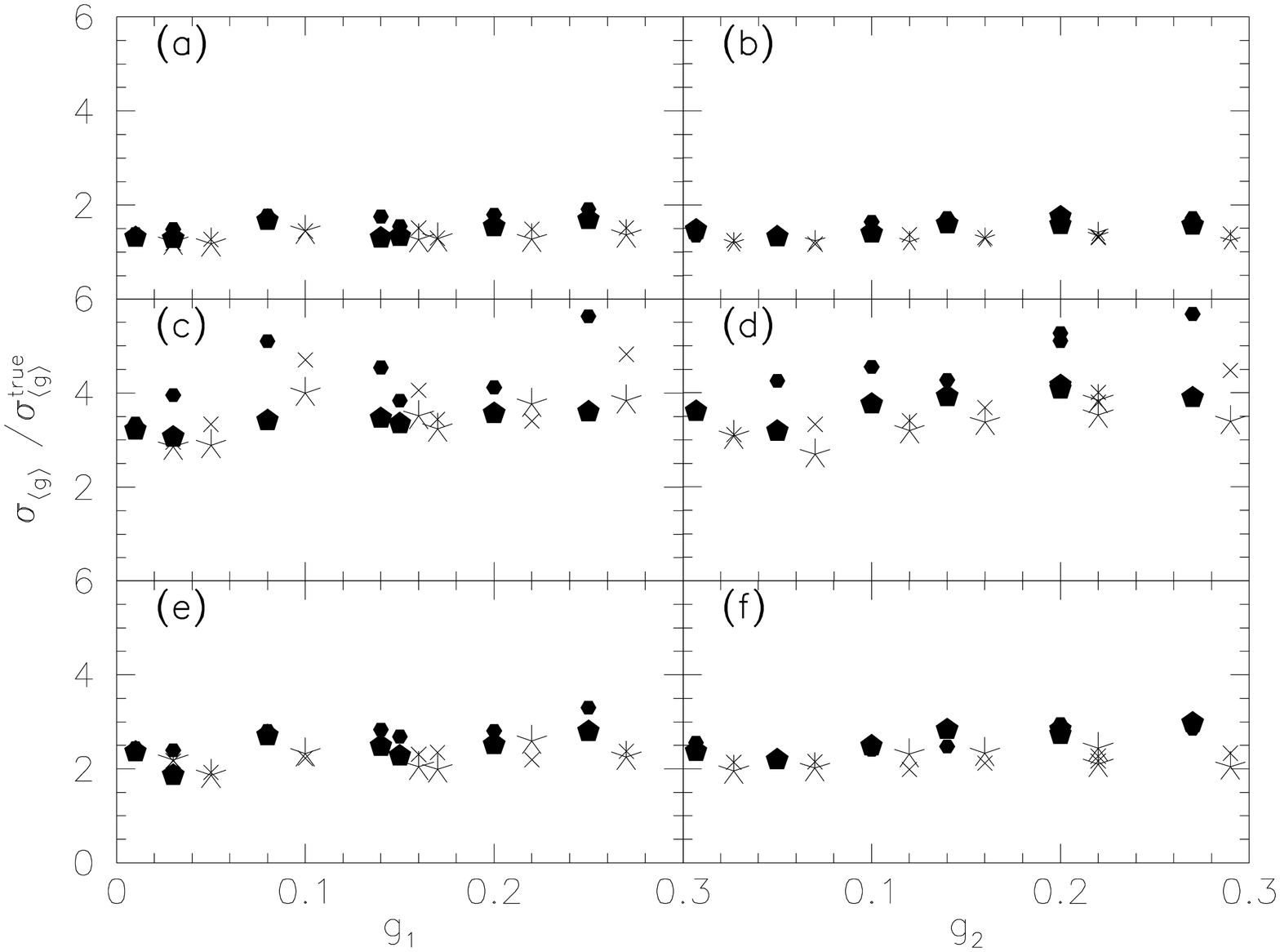,height=0.4\textheight,width=0.9\textwidth}
  \caption{The noise properties of the shear estimations from Fig.
  \ref{measuredvalues}. For high S/N objects, sky noise only has a
  very weak contribution to the shear estimates. It increases the
  uncertainty by a factor 3-5 for low S/N objects and by a factor of about 2-3
  for the whole sample. Symbols for the scalar correction are slightly
  shifted to the right for clarity of the plot. The panels are for the
  same object samples as in Fig. \ref{measuredvalues}.}
  \label{noisevalues}
\end{figure*}

\begin{figure*}[ht]
  \psfig{figure=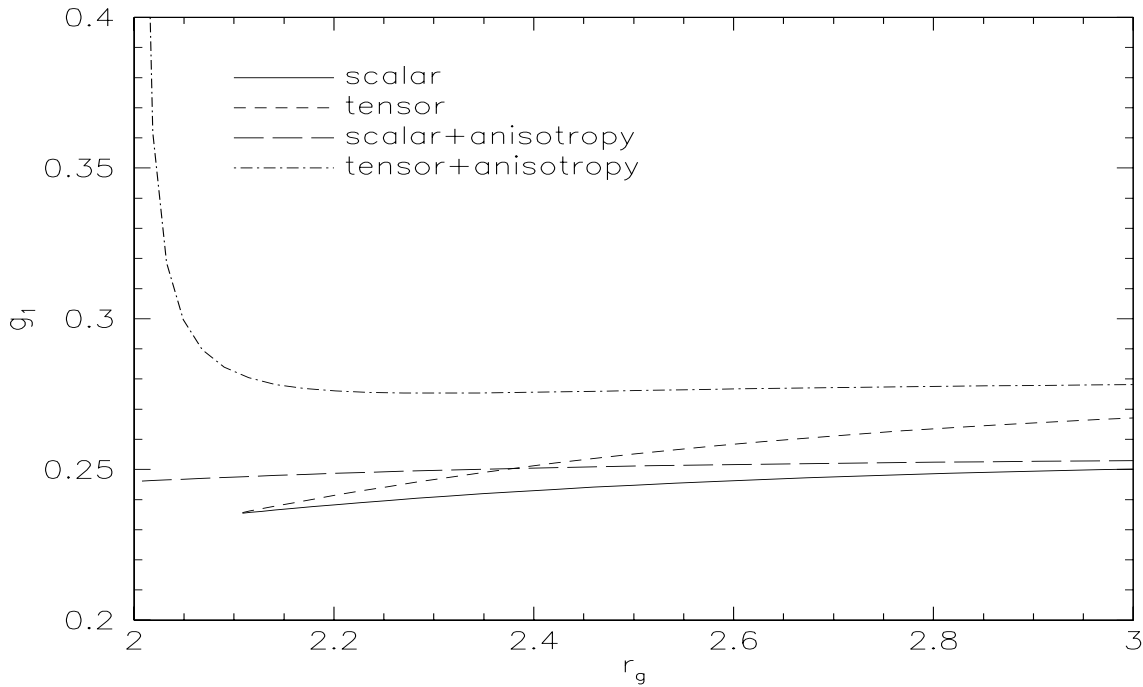,height=0.4\textheight,width=0.9\textwidth}
  \caption{Analytical results for $g_1=0.25$, round intrinsic objects, 
  no PSF anisotropy and a $10\%$
  PSF anisotropy in the $g_1$ direction are shown. The solid and short
  dashed lines are scalar and tensor shear estimate for the PSF
  anisotropy free case, the long dashed and dashed dotted line 
  represent scalar and tensor correction when PSF anisotropy is
  present. The figure confirms that the resulting shear is
  overestimated significantly in the tensor case with PSF anisotropy.}
  \label{rganisofig}
\end{figure*}

To evaluate how accurately we can measure gravitational shear we have
investigated the $g_1$, $g_2$ combinations listed in Table
\ref{summarytable}. We did our measurements first without PSF
anisotropy and then with a stellar axis ratio
$b_{*}/a_{*}=0.9$ with $a_{*}$ oriented such that $q_2^*=0$.  To
obtain the results quoted below, we have calculated averages
$\langle \gvec\rangle$ and errors $\sigma_{\langle \gvec\rangle}$ over
all galaxies in our catalogs.  The results for the noise-free
calculations are shown in Fig. \ref{truevalues}. Note here that we
have also used the same weighting scheme for the noise-free calculations,
although this would not be necessary.
From the plot we see the following trends for the calculations {\it without}
PSF anisotropy:
\begin{itemize}
\item For small input shear values $|\gvec |\le 0.1$ we can clearly recover
the input shear within $\Delta\gvec\approx 0.01$ with the tensor as well as with the
scalar shear estimator. For larger shear values we systematically
underestimate the shear, with a maximal underestimation of $0.02$ 
to $0.03$ for the highest input shear of $0.2-0.25$. We note that
for both shear estimators the underestimation as a function of
input shear can be very well represented by a straight line indicating
that the underestimation is a constant fraction of the input shear
over the whole range of input shears considered here.
This fraction is about $10\%-15\%$ for the scalar and $5\%-10\%$ for
the tensor shear estimate. 
\item We also note that the results are
completely equivalent in both shear components.
\end{itemize}
For the simulations {\it with} PSF anisotropy in the $g_1$ direction
we conclude the following:
\begin{itemize}
\item Considering the $g_2$ component the results are very similar
to the isotropic PSF case. 
\item For the $g_1$ component the scalar correction underestimates the
shear by an amount comparable to the isotropic PSF case. In
contrast, the tensor correction overestimates the input shear in the direction
of the PSF anisotropy. This relative
overestimate is about $20\%$ for the high $|\gvec|$ and can reach up to
$100\%$ for the lowest input shears.
In Fig. \ref{rganisofig} we show analytic results with intrinsically
round objects and an input shear of $g_1=0.25$. These calculations
confirm that for Gaussian profiles the tensor correction overestimates
the final shear in the direction of a PSF anisotropy. 
From our analysis of the SkyMaker simulations in Sect. 
\ref{skymakersimulsec}, where we have
encountered similar problems with the tensor correction, we conclude
that the problems are not connected with the anisotropy correction of
KSB but with the smearing correction.
\end{itemize}
For the calculations including sky noise we have investigated the whole
sample, and two sub-samples with S/N$<2$ and S/N$>2$.
Fig. \ref{measuredvalues} shows that the S/N$>2$ sample
results are similar to the noise-free case, i.e. in the absence of
PSF anisotropy we have very good results with the tensor estimate
(however, there is no underestimation in the tensor case at all in
contrast to the noise-free case; in fact, the underestimation
is lowered to $4\%$ in the noise-free case for the same subsample
of galaxies),
and about a $10\%-15\%$ underestimate for the scalar calculations. 
For the low signal-to-noise objects
(S/N$<2$) the shear is underestimated by about $30\%$ in the scalar
and up to about $50\%$ in the tensor case, where the underestimate
in the tensor case shows an increasing trend for increasing shear. 
The error in the object centre determination
is not the main reason for the underestimate:
for S/N$<2$ we repeat the same calculations using the true object
centres. The results from this test are very comparable to those shown here.
For the whole sample, the scalar correction is better with
a relative underestimate of about $10\%-15\%$ over the tensor one which has an
underestimate of about $20\%-30\%$. If PSF anisotropy is present,
noise-free and noisy cases give the same trends.
Fig. \ref{noisevalues} shows the noise amplitude of the estimators.
We considered the ratio of the shear uncertainty
from noisy objects $\sigma$ and the noise free calculations
$\sigma^{\rm true}$.  For S/N$>2$ pixel noise increases the errors in
the shear estimates by about a factor of 2, for S/N$<2$,
by a factor 5-10, and by a factor of 5 for
the whole sample. 

\begin{figure*}[ht]
  \psfig{figure=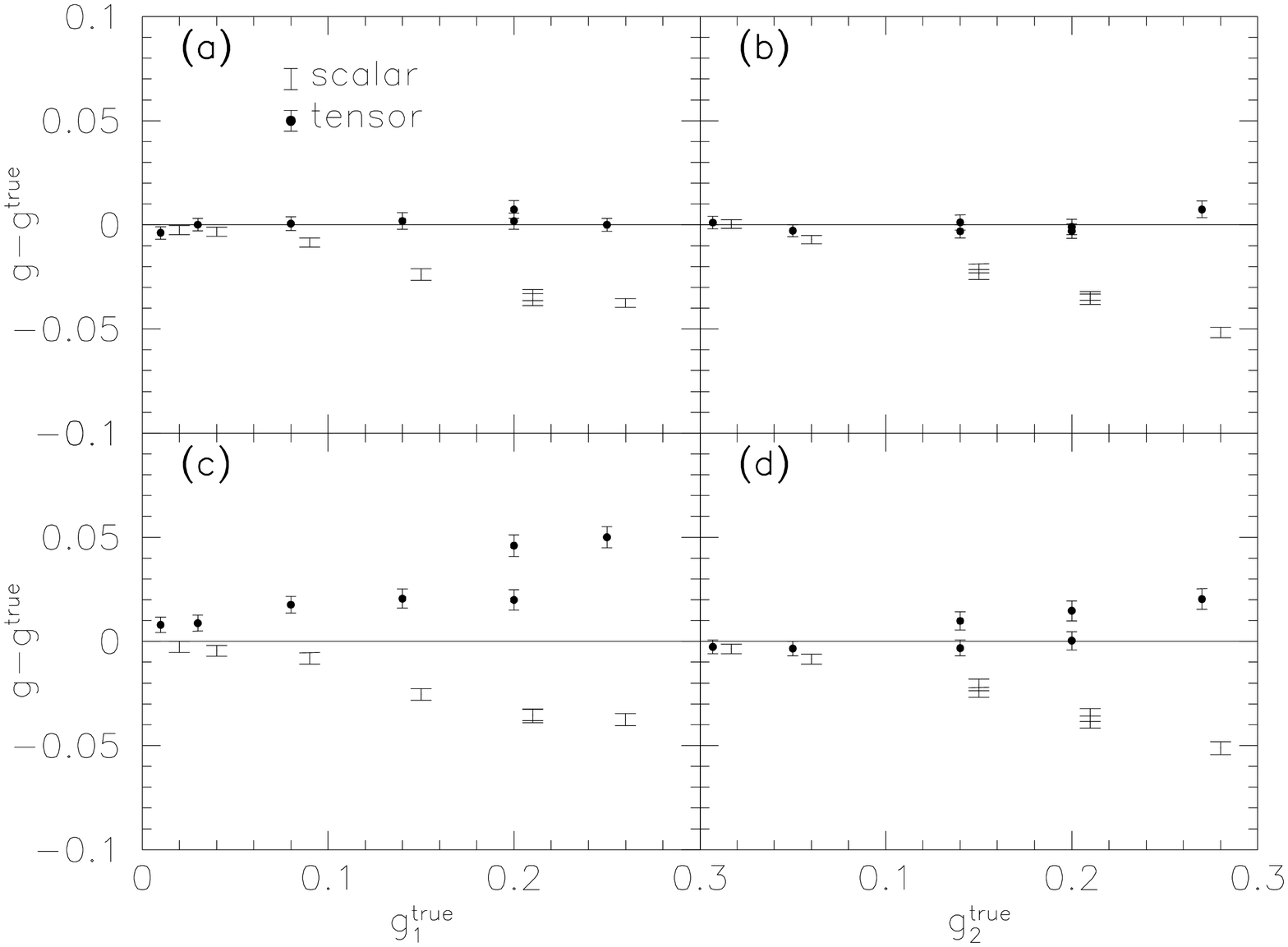,height=0.4\textheight,width=0.9\textwidth}
  \caption{Results for Gaussian profiles put on FITS images and
  analysed with the procedure outlined in Sect. \ref{selectionsec}
  are shown. Panels (a) and (b) are for the PSF anisotropy free case and
  the panels (c) and (d) for a PSF anisotropy of $10\%$ in the $g_1$
  direction. All galaxies were used and no selection according 
  to S/N was done. Dots with errorbars represent measurements with tensor
  estimation, lines with errorbars those with scalar estimation. The
  final estimates were done with about 13500 galaxies for every
  realisation and error bars have a typical size of $1\%$.
  The results are completely comparable to the
  S/N$>2$ case of Fig. \ref{measuredvalues}}
  \label{gaussresfig}
\end{figure*}
\subsection{Summary}
With our proposed way to apply the KSB technique to measure
gravitational shear we come to the following conclusions when dealing
with Gaussian profiles:
\begin{itemize}
\item Our analysis shows that the scalar estimator from eq. 
(\ref{shearestimatebad}) systematically underestimates shear
by about $10\%-15\%$. For low-S/N objects, the underestimate is about
$30\%$. The tensor estimator from eq. (\ref{shearestimategood}) is 
better for high-S/N objects showing no systematic over- or 
underestimation in the final result. This estimator is less stable
for low-S/N objects where there is a systematic underestimation
that can reach more than $50\%$. When including PSF anisotropy, 
the scalar estimator still gives very stable and comparable results
both in the direction of the PSF anisotropy and perpendicular to it.
In contrast, we overestimate the shear in the direction of PSF
anisotropy in the tensor case. We conclude that the scalar estimator
is more stable and conservative. Both estimators show very similar
noise properties.
\item We have shown that smoothing $P^{\rm g}$ does not improve
the final shear estimates over taking raw, noisy values.
\item We can give an objective parameter cut S/N$>2$ for which
we can measure shear with about the same accuracy as with no sky noise.
Our S/N parameter [see eq.(\ref{SNratio})] is very convenient as
we do not have to find the threshold 
for every observation individually (as would be the case for
quantities like m).
\end{itemize}
As a final step we have repeated the analysis done here putting our
Gaussian galaxies and stars on FITS images and analysing them in the
same way as
the SkyMaker simulations described in the next section. In addition
to the steps
presented so far, we first have to perform object detection
and selection for the shear determination. The steps
for this are summarised below in Sect. \ref{selectionsec}.
An important difference with the previous semi-analytical calculations
is the fraction of S/N$<2$ objects. Previously they constituted more
than $60\%$ of all the objects, but for the simulations where objects have
to be detected first,
this fraction is very small, about $5\%-10\%$ in the final catalogs.
It turned out that shear
estimates with the subsample S/N$>2$ and the complete sample
are now similar, so that the small subsample of low-S/N
objects becomes unimportant. Therefore, only the results with
all objects without a cut in S/N will be shown from now on.
Fig. \ref{gaussresfig} displays the results from the simulations.
They are comparable to those obtained
from the S/N$>2$ subsamples in Fig. \ref{measuredvalues}. We also conclude
that the stellar objects detected in the FITS images have sufficient
signal to noise to give reliable estimates for $P^{\rm g}$. As described
above we only used very high S/N stellar objects in the analytic
calculations. 
This last
simulation ensures that our shear analysis pipeline gives
results consistent with analytical prediction, and that it is ready to
be used for the final analysis involving the realistic simulations
done with SkyMaker.
In the next section, we first describe our procedure leading from
the image frames to a galaxy catalog for shear measurements.
Afterwards we will see whether our results are still
valid when considering different galaxy profiles and especially
more realistic PSFs.
\section{Object detection and selection}
In this section we describe a fully automatic procedure to go
from image frames to a final object catalog for shear analysis.
We note that our main intention was to obtain a catalog ensuring
reliable measurements. We demonstrated at the end of the last section
that our analysis of isolated objects in the semi-analytical treatment
and with those detected in FITS images gives very comparable results.
Our procedure contains several conservative rejection criteria for
objects and it is not optimised to make the maximal use of data in
terms of number density of objects. The procedure consists of the
following steps:
\label{selectionsec}
\begin{figure*}[ht]
  \psfig{figure=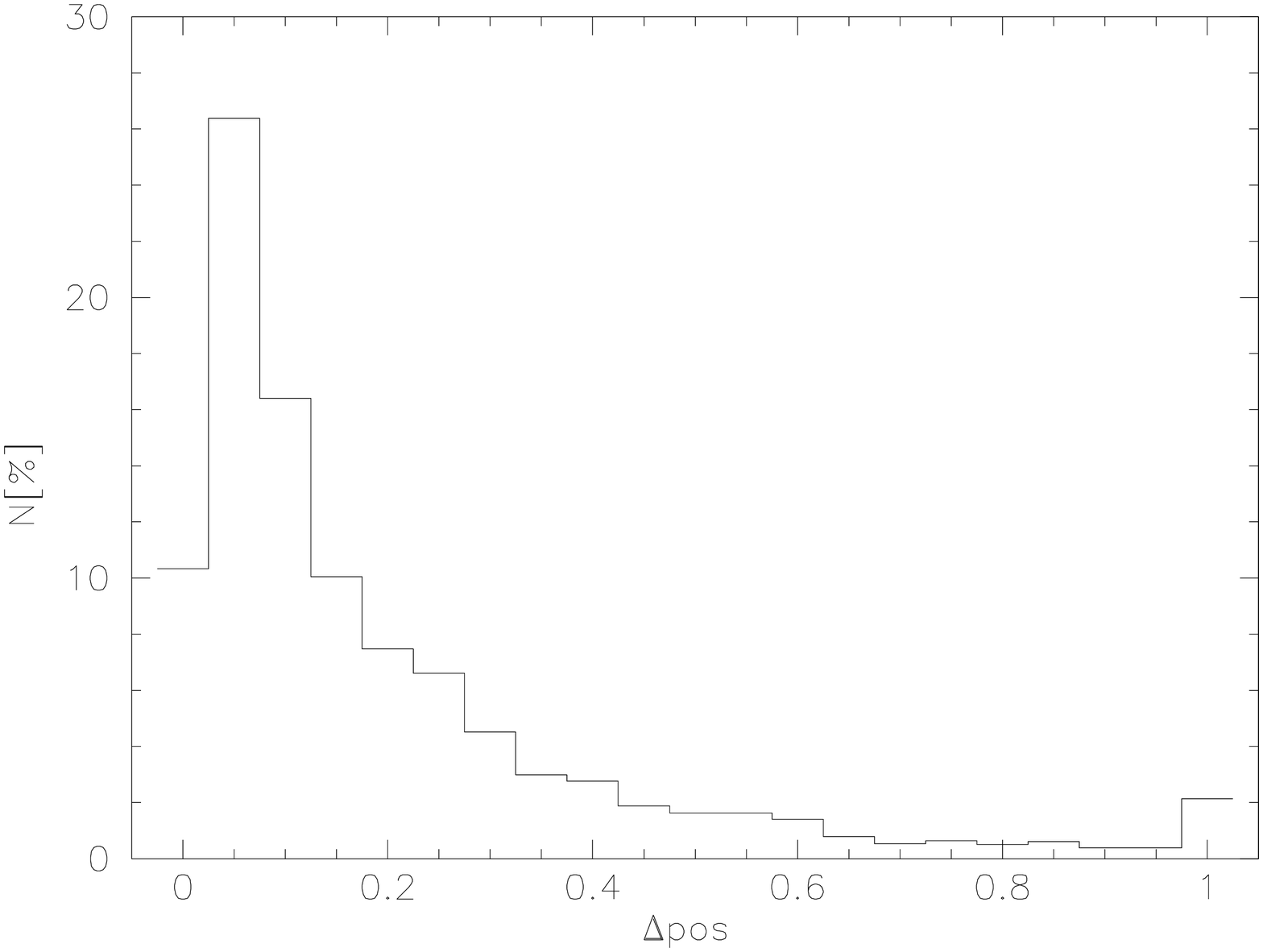,height=0.4\textheight,width=0.9\textwidth}
  \caption{The distribution of the positional differences of the
           `hfindpeaks' position estimator and ours from
           eq. (\ref{goodposeq}). We see a clear peak below 0.1 pixel and so
           we conservatively excluded all objects with a higher difference.
           See the text for more details.}
  \label{posdifffig}
\end{figure*}
\begin{enumerate}
\item Objects were detected with the `hfindpeaks' algorithm from Nick
Kaiser: The image is smoothed with Mexican top hat filters $W_{\rm d}(|\theta|)$
\be 
  W_{\rm d}(|\theta|)=\frac 1{2 r_{\rm g}^2\pi}\left[\exp\left(-\frac{|\theta|^2}{2
  r_{\rm g}^2}\right)- \frac{1}{r_{\rm f}^2} \exp\left(-\frac{|\theta|^2}{2
  r_{\rm f}^2r_{\rm g}^2}\right)\right] 
\ee 
with increasing filter radii $r_{\rm g}$. In
every smoothed image, the peaks are detected and linked with the peaks
found in previous smoothings. Hereby peaks are assumed to belong to
the same object if their positions coincide within $r_{\rm g}$. In this way
a peak trajectory is built up for every potential object.  For every
peak, a signal-to-noise ratio $\nu$ is calculated where the signal is
the peak value in the smoothed image and the noise $\sigma_{\rm d}$ is given
by
\begin{eqnarray}
  \sigma_{\rm d}&=&\sigma_{\rm sky}\sqrt{\int W_{\rm d}^2(\theta){\rm
  d}^2\theta} \nonumber \\
  &=& \sigma_{\rm sky}\sqrt{\frac 1{2\sqrt{\pi}r_{\rm g}}\left(1+\frac
  1{r_{\rm f}^2}-\frac 4{1+r_{\rm f}^2}\right)},
\end{eqnarray}
with $r_{\rm f}=2$. Fig. \ref{snrnufig} shows a comparison
between this signal-to-noise estimate and ours from eq.
(\ref{SNratio}).
\begin{figure*}[ht]
  \psfig{figure=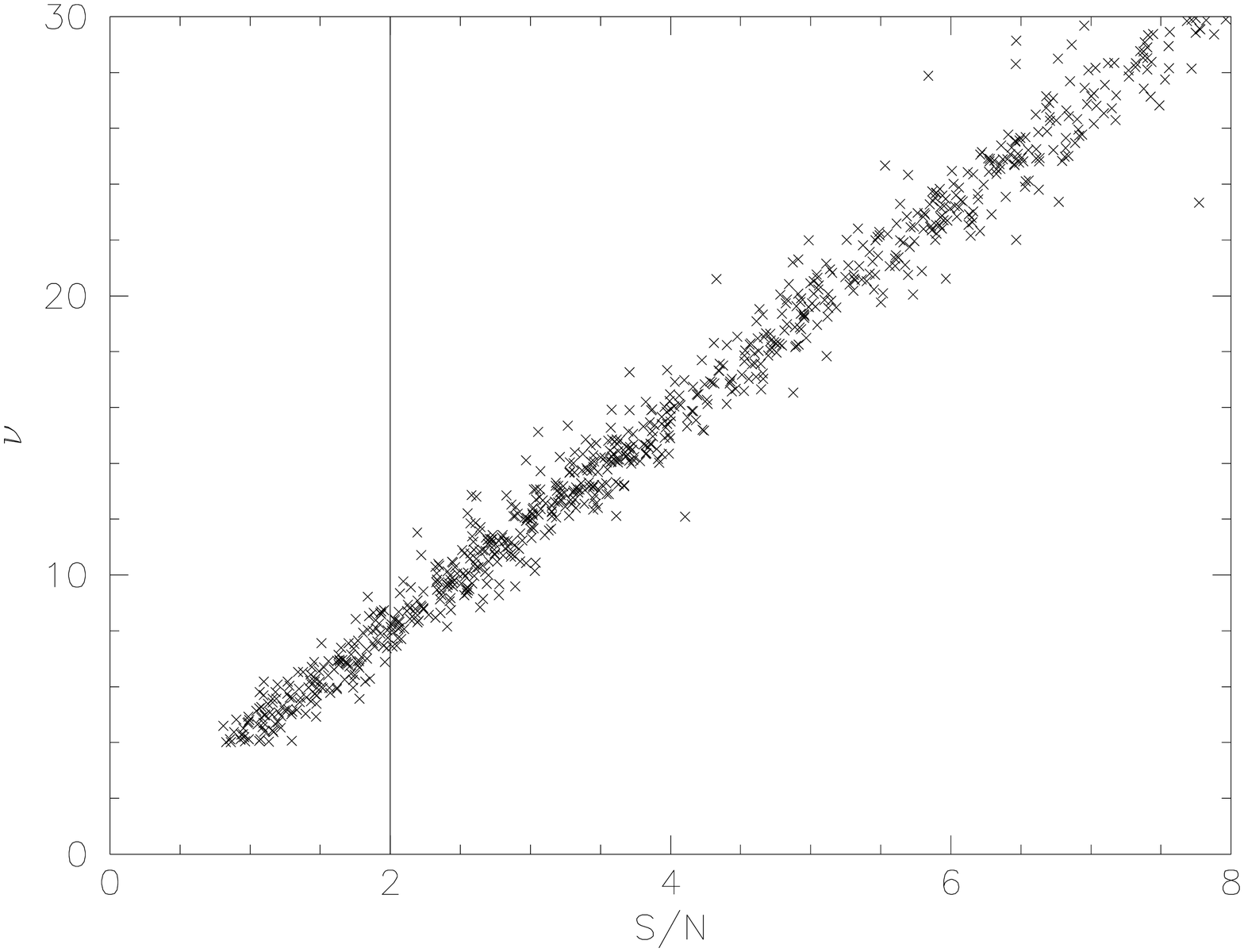,height=0.4\textheight,width=0.9\textwidth}
  \caption{A comparison between the `hfindpeaks' signal-to-noise
  estimate $\nu$ and ours from eq. (\ref{SNratio}). The relation
  is a straight line and our criterion for a good object (S/N$>2$)
  corresponds to $\nu\approx 7$.}
  \label{snrnufig}
\end{figure*}
Afterwards, peaks with the highest
$\nu_{\rm max}$ value for each trajectory identify objects
(peaks with $\nu_{\rm max} < 4$ are immediately rejected). This
size $r_{\rm g}$ is used in all the following analyses and
the pixel centre of the peak position is taken as a starting point
for the object centre determination. To get a more accurate object centre, a
Newton-Raphson step is performed and the $x$ and $y$ pixel-positions 
are corrected from the pixel centre by $\delta_{x}$ and $\delta_{y}$
\be
  \delta_{x}=-\frac{I_{x}^s}{I_{xx}^s}; \;\;\;
  \delta_{y}=-\frac{I_{y}^s}{I_{yy}^s}
\ee
where $I_{x}^s$, $I_{y}^s$ and $I_{xx}^s$, $I_{yy}^s$ are the first and second
derivatives of the smoothed light profiles along $x$ and $y$.
\item From the peaks (objects) found in the previous step, all
relevant quantities (like $P^{\rm sh}$, $P^{\rm sm}$) are 
calculated separately with the original, unsmoothed image.
\item From the catalog generated in the previous step, we first removed
objects with obvious problems during the
analysis process: (1) Objects closer than $3r_{\rm g}$ to the
border of the image; (2) objects where one of the eigenvalues $A^2$,
$B^2$ from the
$Q_{ij}$ tensor was negative.
%
\be
  \label{ellipseparameq}
  {A^2\choose B^2} =\frac{Q_{11}+Q_{22}}2 {+ \choose -}
  \sqrt{\left(\frac{Q_{11}-Q_{22}}2\right)^2+4Q_{12}^2};
\ee
(3) the object has a total negative flux. (4) Fig. \ref{posdifffig}
shows the distribution of the difference between the centroid estimates
of `hfindpeaks' and ours from eq. (\ref{goodposeq}). All
objects with a difference larger than 0.1 pixel were excluded.
We found that this is an efficient way to reject blended objects,
since one centroid estimator is based on the smoothed and the other on
the unsmoothed image.
\label{sortoutstep}
\item Objects with a neighbour within $3 r_{\rm g}$ are rejected.
\item Stars are preselected using the star branch of the $r_{\rm h}-$m diagram 
and polynomials for the two components of $\qvec^*$ are calculated in the following way:
a preliminary fit is done for $q^*_1$ and $q^*_2$ using a $\chi^2$ minimisation
\be
  \chi^2_j = \sum_{i=1}^{N^{*}}(q^*_j(x_i,y_i)-p_j(a_k,x_i,y_i))^2,
\ee
where $N^{*}$ is the number of preselected stars, $j=1,2$,
$x_i, y_i$ are the positions of the stars and $p_j$ are
two-dimensional, second order
polynomials with six unknown parameters $a_k$. After determination
of $a_k$ we calculated the expected error $\sigma_{q_j}$ for every
$q^*_j$ by
\be 
  \sigma_{q^*_j}=\sqrt{\frac{{\chi^2_j}_{\rm min}}{N^{*}-1}}, 
\ee 
where
${\chi^2_j}_{\rm min}$ is the minimum of $\chi^2_j$ at the fitted parameters
$a_k$. Stars which deviate at more than $1\sigma_{q_j}$  in any of the two
components $q_1$ or $q_2$ are rejected, and
the fit is repeated for the final polynomials.
\item The final sample of stars is reprocessed nine times
with filter scales $r_i=1..9$, in order to match all the possible galaxy
sizes.
After each processing, objects with problems according to step
(\ref{sortoutstep}) are still rejected. From the remaining stars
we calculated the mean of $\langle {\rm tr}[P^{* \rm sh}]/{\rm tr}[P^{* \rm sm}]\rangle$ 
and we used this value for the PSF correction.
\item We now have all the quantities in hand to
calculate scalar and tensor shear estimates for every objects as
described in Sect. \ref{applicationref}. Hereby we considered
only objects as galaxies whose half light radius was larger than
the stellar locus (see Fig. \ref{rhmagfig}).
\item Weights for the tensor and scalar $\gvec$ are calculated, 
which is an important ingredient for the two shear
estimators as described in Sect. \ref{applicationref}.
With the whole procedure we end up with a number density of about
30 galaxies per sq. arcmin.
\end{enumerate}
\section{The SkyMaker simulations}
\label{skymakersimulsec}
\begin{figure*}[ht]
  \psfig{figure=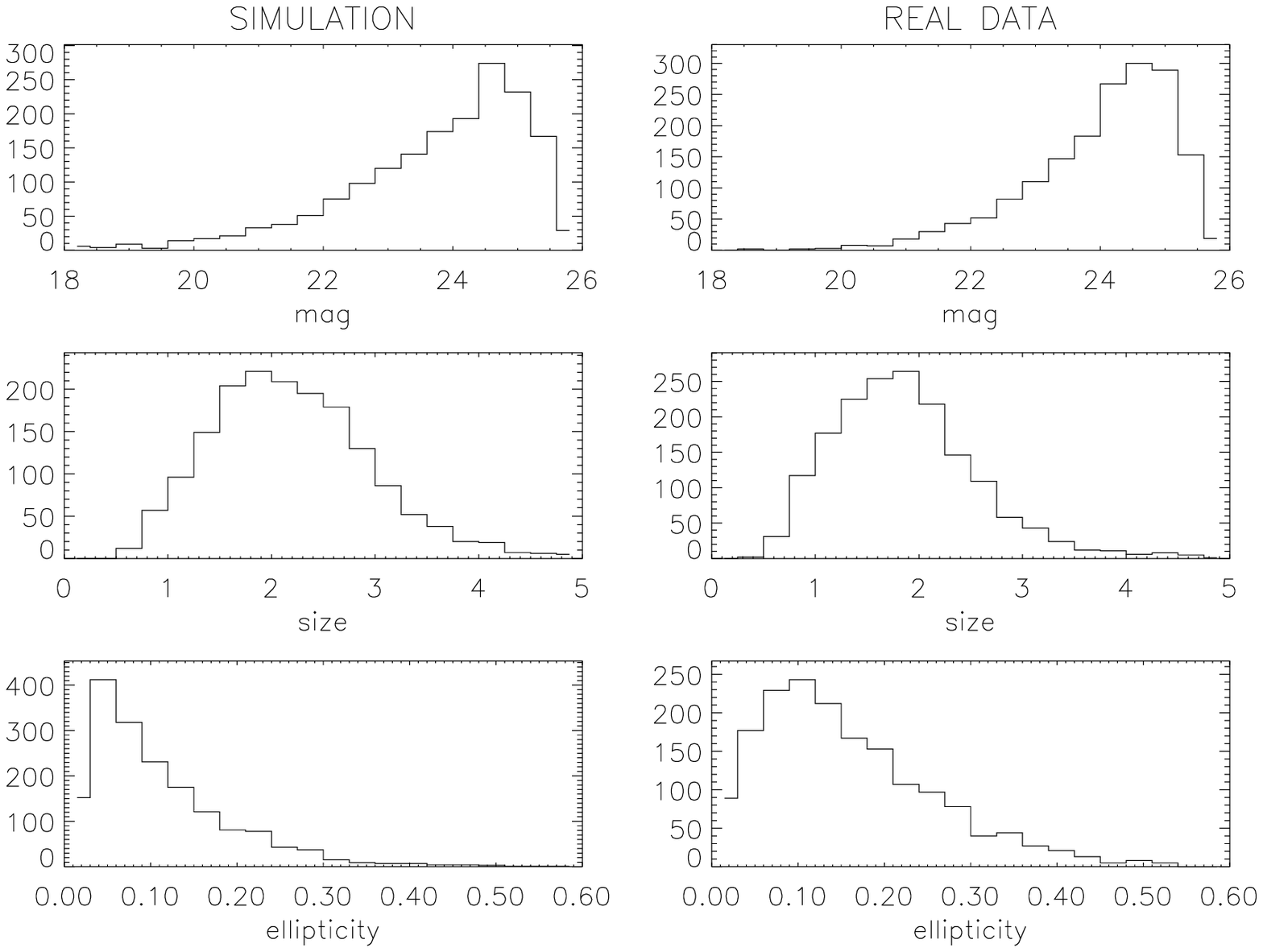,width=1.0\textwidth}
  \caption{A comparison of galaxy properties in one of our SkyMaker images
  with real data in one of our $I$ band CFHT fields. Both images had a
  seeing of $0\arcsecf 7$ and both corresponded to 12000 sec. exposures.
  Objects have been detected and the quantities have been calculated 
  with SExtractor.
  The two upper panels show a comparison of the magnitude
  distribution, the two middle panels of the object size defined as
  $\sqrt{AB}$, where $A$ and $B$ are defined as in
  eq. (\ref{ellipseparameq}) and the lower panels of the ellipticity
  parameter $\epsilon=(1-B/A)/(1+B/A)$. The only noticeable strong
  difference is that the ellipticity distribution in the real data
  is slightly broader. The shift of the peak in this distribution is
  caused by a strong PSF anisotropy in the CFHT data.}
  \label{skyrealcompfig}
\end{figure*}
\subsection{The image characteristics of our SkyMaker simulations}
\label{skysimulsec}
For every combination of the shear and PSF, ten SkyMaker images
were produced and analysed. Each image has
the characteristics of a 12000 sec. $I$ band exposure taken at a telescope
with a 3.5m primary and a 1m secondary mirror, and has a
dimension of $2048\times 2048$ pixels with a scale of 0\arcsecf 206 per 
pixel. The atmospheric seeing in every image was 0\arcsecf 7. 
For every image we detected our objects and analysed 
them as described above. A comparison of magnitude distribution,
object size and ellipticity distribution between detected objects in
one of the images and a 12000 sec. $I$ band CFHT image
having a measured seeing of 0\arcsecf 7 is shown in Fig. \ref{skyrealcompfig}.
A description how the PSF is constructed in SkyMaker is given in
the Appendix.
\subsection{Results of the SkyMaker simulations}
\begin{figure*}[t]
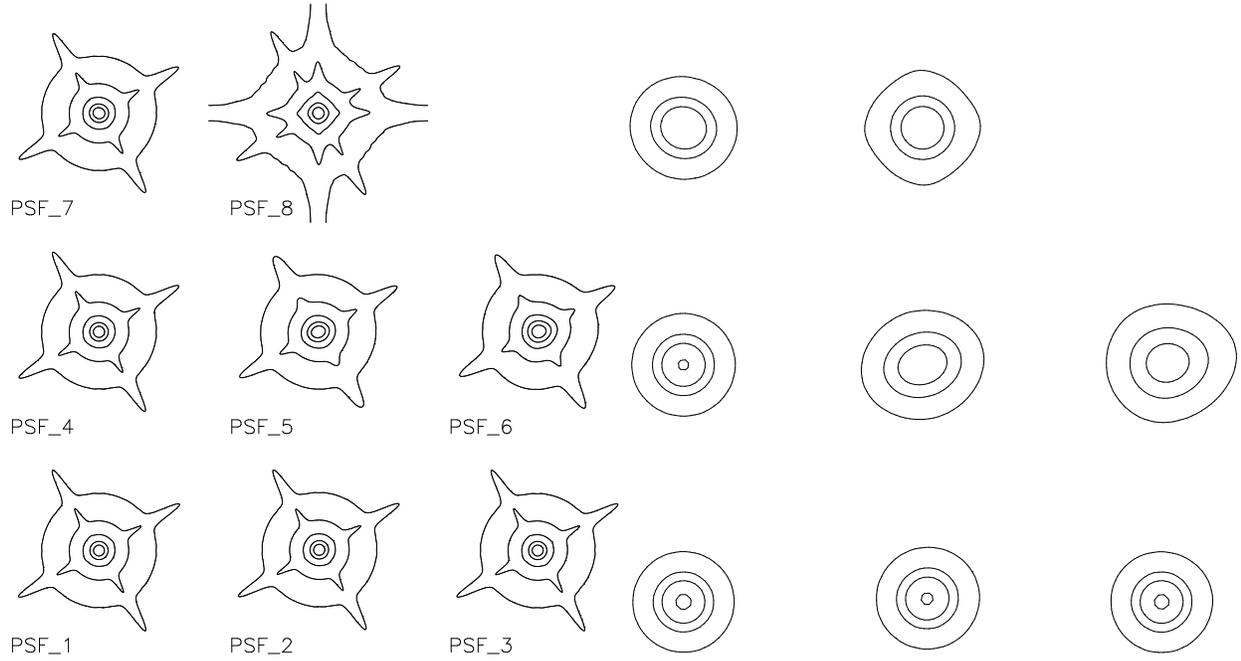

  \begin{center}	
  \begin{minipage}[t]{0.45\textwidth}
    \psfig{figure=H2239_13.ps,width=\hsize,angle=-90}
  \end{minipage}
  \begin{minipage}[t]{0.45\textwidth}
    \psfig{figure=H2239_14.ps,width=\hsize,angle=-90}
  \end{minipage}
  \end{center}
  \caption{The figure shows the outer part (left) and the core (right)
  of the PSFs used in the SkyMaker simulations. All the PSFs contain
  four spider arms produced by the support of the secondary mirror.
  From left to right and from bottom to top the PSFs contain the
  defects (For the tracking errors, the first value in the
  bracket gives the rms drift in the direction of the second value.
  The third value is the rms of a Jitter in the orthogonal direction.
  For the aberrations the ``{\tt d80}'' diameters introduced are quoted
  (see the appendix for a detailed description of tracking errors and
  aberrations). We note that the ``{\tt d80}'' diameters
  of the individual aberrations do not add quadratically as the aberration
  polynomials are not orthogonal.):   
  (1) None; 
  (2) tracking error (0.0'', 97.0$^{\circ}$, 0.05''),
  defocusing (0.05''), coma (0.26''); 
  (3) tracking error (0.05'', 0.0$^{\circ}$, 0.01''), 
  defocusing (0.07'') and triangular aberration (0.015'');
  (4) tracking error (0.02'', 34.0$^{\circ}$, 0.02''),
  defocusing (0.1'') and spherical aberration (0.3'');
  (5) tracking error (0.3'', 31.0$^{\circ}$, 0.2''),
  defocusing (0.3'') and astigmatism (0.72'');
  (6) tracking error (0.3'', 62.0$^{\circ}$, 0.3''),
  defocusing (0.3''), coma (0.31'') and triangular aberration
  (0.72'');
  (7) tracking error (0.5'', -20.0$^{\circ}$, 0.11''),
  defocusing (0.07'') and triangular aberration (0.122'')
  (8) defocusing (0.05'') and quadratic aberration (0.61'')}
  \label{psfsfig}	
\end{figure*}

\begin{figure*}[ht]
  \psfig{figure=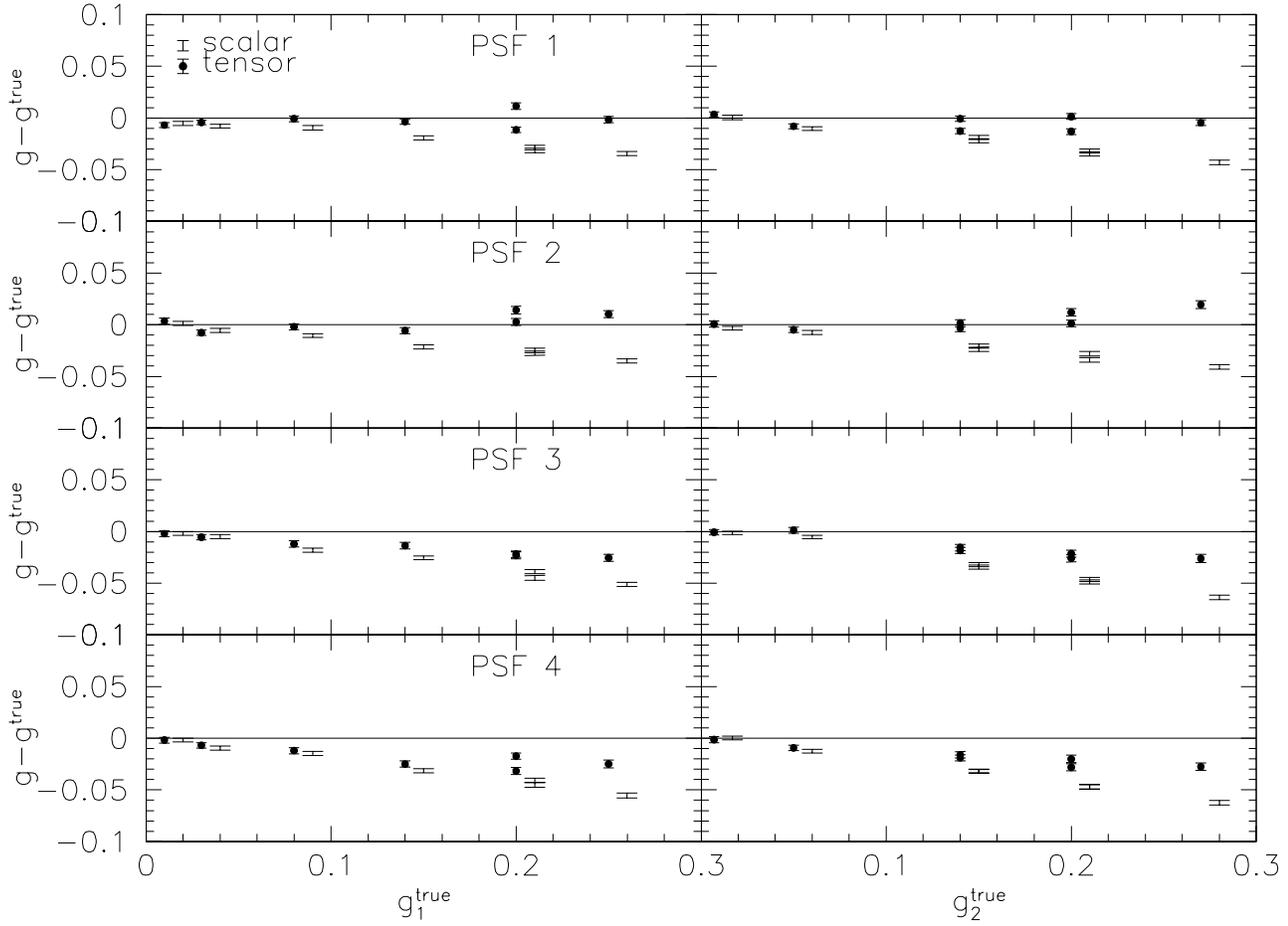,width=1.0\textwidth}
  \caption{From top to bottom the figure shows SkyMaker results for
  the first four PSFs of Fig. \ref{psfsfig}. The objects from the 
  ten images for every shear/PSF combination were pasted together and
  the means and errorbars shown here were obtained from this pasted 
  catalog.
  }
  \label{skymakerres1fig}
\end{figure*}

\begin{figure*}[ht]
  \psfig{figure=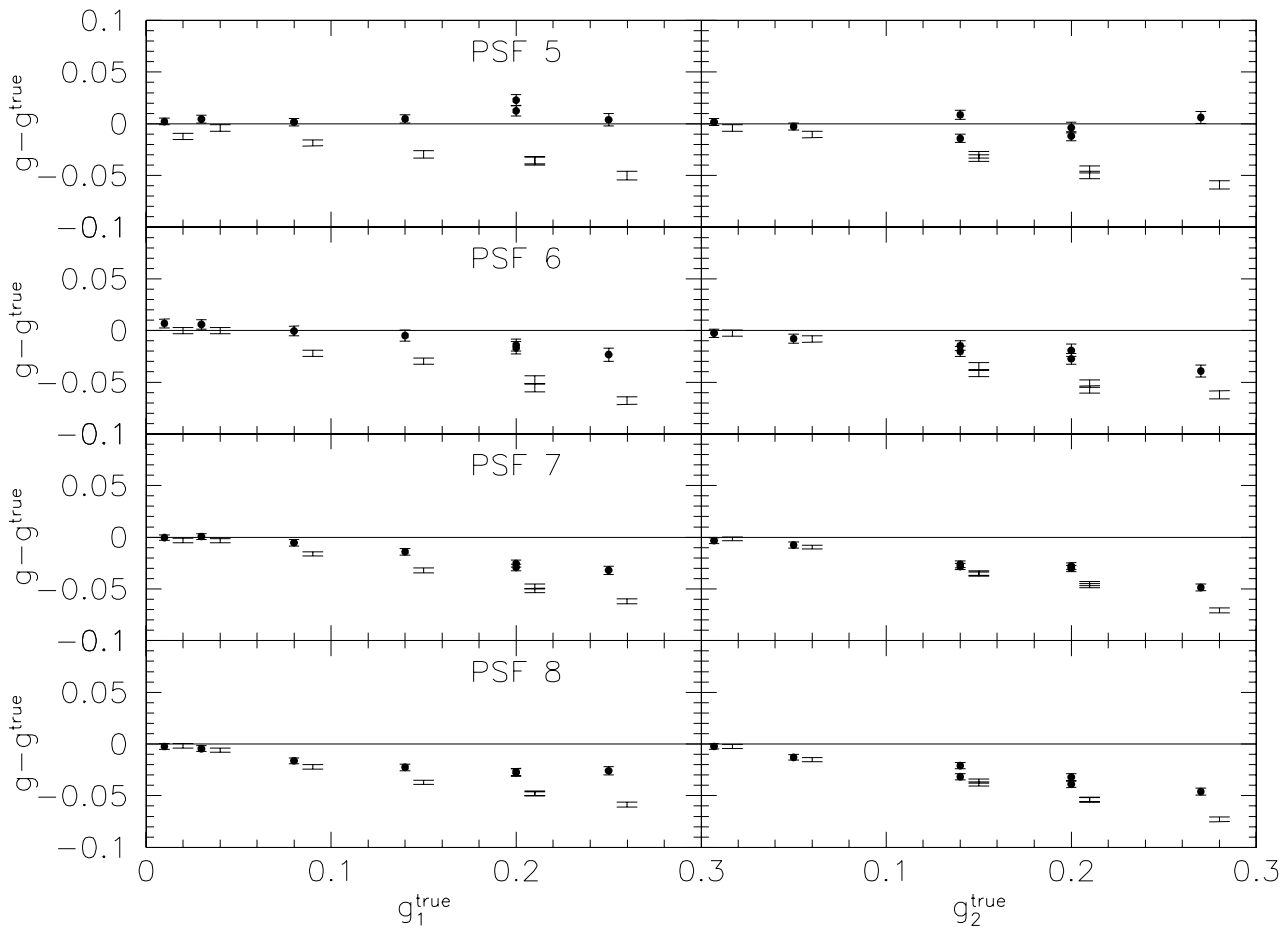,width=1.0\textwidth}
  \caption{The same as Fig. \ref{skymakerres1fig} for the second four
  PSFs of Fig. \ref{psfsfig}.}
  \label{skymakerres2fig}
\end{figure*}

\begin{figure*}[ht]
  \psfig{figure=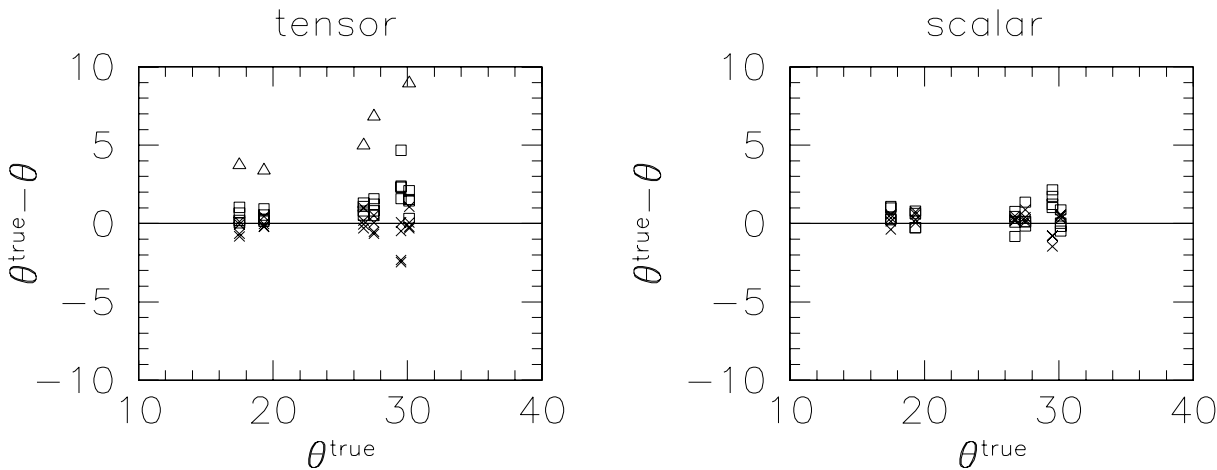,width=0.95\textwidth}
  \caption{The figure shows the difference ($\theta^{\rm
  true}-\theta$) of the recovered position angles vs.
  the position angles of the input shear. Crosses correspond to the results from Fig.
  \ref{skymakerres1fig} and diamonds to those from
  Fig. \ref{skymakerres2fig}. Triangles are for shear values where
  no PSF anisotropy correction has been done (PSF\_4). In this plot
  we omitted the very weak shear case ($g_1=0.01$, $g_2=0.007$) as the
  position angle is badly determined here.}
  \label{anglesfig}
\end{figure*}

Fig. \ref{psfsfig} shows contours of the outer and core parts of the
PSFs we have used in our simulations. The outer parts of
the profiles all look the same (except for the quadratic PSF) and they
differ mostly in the cores. The anisotropies caused by these PSFs are
given in Table \ref{psfanisotab}. 
\begin{table}
\begin{center}
\begin{tabular}{|c|c|c|c|c|c|c|c|c|}
\hline
& PSF\_1 & PSF\_2 & PSF\_3 & PSF\_4 \\
\hline
$\chi_1$ & 0.00  & 0.00 & 0.00 & 0.00 \\ 
\hline
$\chi_2$ & 0.00 & 0.01 & 0.00 & 0.00 \\
\hline
\end{tabular}
\begin{tabular}{|c|c|c|c|c|c|c|c|c|}
\hline
& PSF\_5 & PSF\_6 & PSF\_7 & PSF\_8 \\
\hline
$\chi_1$ & 0.1  & 0.06  & 0.04 & 0.00\\ 
\hline
$\chi_2$ & 0.07 & 0.03 & -0.03 & 0.00\\
\hline
\end{tabular}
\end{center}
\caption{\label{psfanisotab} The anisotropies of the PSFs described
in Fig. \ref{psfsfig}. The values quoted are the raw ellipticity
measurements from KSB where the $r_{\rm g}$ from the stars was used
as filter scale.}
\end{table}
For every PSF we have analysed
seven sets of images with the shear combinations given in Table
\ref{summary2table} where a few sets are slightly different from
those in Table \ref{summarytable}.
\begin{table}
\begin{center}
\begin{tabular}{|c|c|c|c|c|c|c|c|}
\hline
$g_1$ & 0.01  & 0.03 & 0.25 & 0.08 & 0.2  & 0.2  & 0.14 \\ 
\hline
$g_2$ & 0.007 & 0.05 & 0.2  & 0.14 & 0.27 & 0.14 & 0.2 \\
\hline
\end{tabular}
\end{center}
\caption{\label{summary2table} The shear combinations we investigated
with our SkyMaker simulations.}
\end{table}
The results for our SkyMaker simulations are
shown in Figs. \ref{skymakerres1fig}, \ref{skymakerres2fig},
\ref{anglesfig} and Table \ref{skysummarytable}. 
As for the semi-analytical and simulated Gaussian profiles
we now discuss the accuracy of the shear estimators for the different
shear amplitudes and PSF anisotropies:
\begin{itemize}
\item In the case of no PSF anisotropy at all (PSF\_1), the tensor correction
gives the correct shear on the whole range of $|\gvec |=0.012$ up to
$|\gvec |=0.32$. The scalar correction underestimates the shear
relatively by about
$10\%-15\%$. The $1\sigma$ error bars are about at $\pm 0.005$.
\item In the presence of PSF anisotropy, the tensor correction can
over- or underestimate the true shear while it is always underestimated
in the scalar case. In the sense of deviation from the true value,
the tensor case always gives the better result. The relative
underestimates with the scalar correction can reach up to $30\%$ in
the worst case. As the anisotropy of stars is perfectly corrected
by our polynomials for $p$, we investigated whether there is
nevertheless an anisotropy residual in the galaxies. For this we
compared the position angle from the input shear with the recovered
one in Fig. \ref{anglesfig}. There we see that in the scalar case the
position angle is nearly perfectly recovered. This means that
the PSF anisotropy is corrected very well but the calibration 
factor $P^{\rm g}$ is too low. In the tensor case the recovery
of the position angle is not as good as in the scalar case but also
very acceptable. The reason for the over- or underestimates of the
magnitude of the shear is probably that the elements of $P^{\rm g}$ 
that are used for the isotropic correction should be calculated
from profiles that are already corrected for PSF anisotropy defects.
\item From Table \ref{skysummarytable} we conclude that also for all
the SkyMaker simulations, the over- or underestimates
are nearly a constant fraction of the input shear. For the tensor
estimator we can typically recover the input shear to an accuracy between $10\%-
15\%$. 
This means that we can measure a shear below $0.1$ nearly within
$0.01$, the accuracy required for accurate measurements of the cosmic
shear on scales $\leq 10\arcminf 0$.
\item For all the SkyMaker simulations the recovery of the shear is
better than in the anisotropic case with Gaussian profiles (see
Fig. \ref{gaussresfig}). It seems that the more realistic SkyMaker
PSF profiles better reflect the assumptions in the KSB algorithm
than Gaussian profiles,
namely that PSF anisotropy mainly comes from the central core. 
While a Gaussian has its anisotropy on all
scales, Fig. \ref{psfsfig} shows that the anisotropy details of
the SkyMaker PSFs lie in the core indeed.
\item Regarding the issue of scalar vs. tensor correction, we can
conclude that, although the scalar case always underestimates
the true shear, it provides the more conservative 
answer and it should be used when one
is interested mainly in the position angle of objects while the
tensor should given preference when the amplitude of the shear
is important.
\end{itemize}
\section{Conclusions and Outlook}
In this paper we have investigated how well we can recover weak
gravitational shear with realistic simulations for ground-based
observations. With respect to the main motivation for this work, 
to test the
reliability of our recent detection of cosmic shear, the results
are very encouraging. In all our simulations we could recover weak
shear up to $|\gvec |=0.1$ with an accuracy of $0.01$ or better and we
can significantly exclude the detection of a $|\gammavec |=0.04-0.05$ false signal
based purely on uncorrected PSF effects. 
Although the PSFs we have investigated do certainly not cover all
possible defects mimicking a lens signal, also other independent
studies like Hoekstra et al. (1998), who looked at diffraction-limited
model PSFs for the HST or Bacon et al. (2000b), indicate
that the KSB algorithm works better than one could expect looking
at its assumptions.
We showed that for the PSFs investigated here, the anisotropy 
correction proposed in the original KSB
work is working very well, but that there are problems with the
isotropy correction if an anisotropy is present. The reason
for this is probably that in such a case, the boost factor $P^{\rm g}$ is
calculated with the wrong, not anisotropy-corrected, galaxy profile.
Moreover, we have presented a fully automatic procedure leading from the image
frames to an object catalog for reliable shear measurements. 
Despite these
encouraging results, the advent of new wide-field imaging 
mosaic cameras in particular, and the associated data flow brings additional
technical problems if we want to build up a fully automatic data
processing pipeline. Some of these problems that we could not address
in the scope of this work, but leave to a future publication are:
\begin{enumerate}
\item In wide-field imaging especially, many images suffer from
very bright stars causing blooming effects, strong stray light and
reflections.
So far, we usually have marked out affected image regions by hand, but
it should be investigated whether there are automatic ways of dealing
with it.
\item We usually coadd several single exposures to get a final
image. These frames typically have slightly different seeing disks.
In the past we often simply did not use images that had the worst
seeing. We have to investigate what effects the coaddition of images
with different PSF properties has on the final shear measurement result.
\item As a consequence of the last point, weak-lensing observations 
have mostly been done with a very compact dither pattern so far.
This has the advantage of allowing a simple coaddition of the images
with integer pixel shifts, and the effects of optical image distortions
do not need to be taken into account during the image prereduction.
This first preserves uncorrelated noise in the image pixels, second
it preserves the smoothness of PSF anisotropy on the scale of single
chips, and third it brings the practical advantage that we can still
deal with single chips instead of very large images. Nevertheless,
we think that this approach is not ideal. On the one hand accurate
astrometry and photometry is done most easily if information from
objects in the overlap between different chips is at hand. 
The afore mentioned advantages may turn into problems
in later analysis when the gaps between single chips lead
to severe border effects, e.g. when we try to search for filaments
between galaxy clusters (see Kaiser et al. 1999).
So we will investigate whether we can overcome the effects of
correlated
pixel noise caused by remapping for optical distortions
and discontinuities of the PSF anisotropy in the final coadded images.
\end{enumerate}
\begin{landscape}

\begin{table}
\begin{center}
\begin{tabular}{|c|c|c||c|c||c|c||c|c||c|c||c|c||c|c||c|c|}
\multicolumn{17}{c}{\rule[-3mm]{0mm}{8mm}\bf Results of the SkyMaker simulations} \\
\hline
 & $g_1$ & $g_2$ & $g_1$ & $g_2$ & $g_1$ & $g_2$ & $g_1$ & $g_2$ &
 $g_1$ & $g_2$ & $g_1$ & $g_2$ & $g_1$ & $g_2$ & objects & $r_h$ \\
input shear & 0.01 & 0.007 & 0.03 & 0.05 & 0.25 & 0.2 & 0.08 & 0.14 & 0.2 & 0.27
 & 0.2 & 0.14 & 0.14 & 0.2 & per $(1\arcmin)^2$ & (stars) \\
\hline
PSF\_1 scalar & -52.93 & 7.41 & -26.20 & -20.56 & -13.79 & -17.44 & -12.03 & -15.82 & -14.31 & -15.92 & -15.66 &-13.41 & -13.78 & -16.01& & \\
PSF\_1 tensor & -68.73 & 47.70 & -13.98 & -16.13 & -0.67 & 0.70 & -0.97 & -9.03 & 5.72 & -1.69 & -5.75 &-0.51 & -2.61 & -6.56& \raisebox{1.0ex}[-1.0ex]{40} & \raisebox{1.0ex}[-1.0ex]{$\approx 1.95$}\\
\hline
PSF\_2 scalar & 11.39 & -46.60 & -18.62 & -14.96 & -14.01 & -14.08 & -13.10 & -16.95 & -13.77 & -15.09 & -12.50 &-14.68 & -15.36 & -16.98& &\\
PSF\_2 tensor & 33.70 & 8.63 & -26.04 & -9.70 & 4.09 & 5.94 & -2.51 & -2.47 & 7.09 & 7.23 & 1.36 &1.08 & -4.08 & 0.65& \raisebox{1.0ex}[-1.0ex]{28} & \raisebox{1.0ex}[-1.0ex]{$\approx 2.15$}\\
\hline
PSF\_3 scalar & -20.31 & -18.57 & -16.40 & -10.96 & -20.50 & -24.39 & -22.80 & -24.55 & -22.43 & -23.62 & -19.42 &-22.81 & -18.19 & -23.21& & \\
PSF\_3 tensor & -21.30 & -10.27 & -18.62 & 2.43 & -10.20 & -12.95 & -15.13 & -13.14 & -11.52 & -9.69 & -11.01 &-11.09 & -9.80 & -10.78& \raisebox{1.0ex}[-1.0ex]{35} & \raisebox{1.0ex}[-1.0ex]{$\approx 1.95$}\\
\hline
PSF\_4 scalar & -16.54 & 3.09 & -32.02 & -25.30 & -22.12 & -23.36 & -18.46 & -22.62 & -20.34 & -23.12 & -22.56 &-22.94 & -22.48 & -23.59& & \\
PSF\_4 tensor & -18.59 & -18.81 & -22.81 & -18.60 & -9.97 & -10.03 & -14.99 & -13.55 & -8.63 & -10.18 & -15.84 &-11.43 & -17.88 & -14.02& \raisebox{1.0ex}[-1.0ex]{33} & \raisebox{1.0ex}[-1.0ex]{$\approx 2.05$}\\
\hline
PSF\_5 scalar & -121.00 & -55.93 & -13.59 & -20.69 & -20.11 & -24.76 &
-23.08 & -21.41 & -17.98 & -21.92 & -17.73 & -23.75 & -21.24 & -22.06& & \\
PSF\_5 tensor & 21.60 & 25.20 & 14.79 & -5.56 & 1.54 & -1.89 & 1.95 & -10.15 & 11.45 & 2.29 & 6.21 &6.18 & 3.36 & -5.98& \raisebox{1.0ex}[-1.0ex]{21} & \raisebox{1.0ex}[-1.0ex]{$\approx 2.75$}\\
\hline
PSF\_6 scalar & 0.20 & -35.93 & -1.00 & -16.40 & -27.17 & -25.75 & -27.62 & -24.74 & -27.76 & -23.07 & -23.80 &-29.63 & -21.22 & -28.54& & \\
PSF\_6 tensor & 68.21 & -37.49 & 19.49 & -15.74 & -9.36 & -9.60 & -0.57 & -10.46 & -7.10 & -14.53 & -8.41 &-14.44 & -3.47 & -13.71& \raisebox{1.0ex}[-1.0ex]{20} & \raisebox{1.0ex}[-1.0ex]{$\approx 3.05$}\\
\hline
PSF\_7 scalar & -31.28 & -20.51 & -11.07 & -19.00 & -24.77 & -23.41 & -20.16 & -25.42 & -25.79 & -26.21 & -23.81 &-24.70 & -22.89 & -22.43& & \\
PSF\_7 tensor & -2.78 & -47.23 & 2.29 & -15.14 & -12.80 & -13.94 & -6.62 & -20.22 & -12.74 & -17.98 & -14.60 &-18.66 & -10.03 & -15.06& \raisebox{1.0ex}[-1.0ex]{32} & \raisebox{1.0ex}[-1.0ex]{$\approx 2.25$}\\
\hline
PSF\_8 scalar & -17.09 & -33.79 & -19.97 & -30.31 & -23.48 & -26.78 & -27.67 & -27.56 & -24.14 & -27.00 & -23.77 &-25.73 & -26.45 & -27.02& & \\
PSF\_8 tensor & -24.40 & -33.53 & -14.92 & -25.71 & -10.36 & -16.04 & -20.36 & -22.62 & -13.71 & -17.06 & -13.57 &-14.93 & -16.15 & -19.44& \raisebox{1.0ex}[-1.0ex]{30}& \raisebox{1.0ex}[-1.0ex]{$\approx 2.25$}\\
\hline
\end{tabular}
\end{center}
\caption{\label{skysummarytable} The table gives a summary of the
results from the SkyMaker simulations performed. Given is the relative
error $(g_{1,2}-g_{1,2}^{\rm true})/g_{1,2}^{\rm true}$ in percent. 
We note that this error is always a fairly constant fraction over the whole
range of input shears (excluding the very weak shear case). The mean number of used
objects per sq. arcmin and the measured stellar $r_h$ are from the catalogs of the
$g_1=0.01$, $g_2=0.007$ input shear.}
\end{table}

\end{landscape}
\section*{Acknowledgements}
We thank Lindsay King, Matthias Bartelmann and the anonymous referee for helpful discussions
and suggestions. T.E. thanks CITA and IAP for hospitality where part of this work
has been done, and L.V.W. thanks MPA for hospitality where it has been initiated.
We thank Henk Hoekstra for a first introduction
to KSB and for providing his codes. We also used publicly available
codes from Erik Deul, Nick Kaiser, David Mount and Sunil Arya without
which this work could not have been done. We thank the TERAPIX data
center in Paris and the ESO Imaging Survey team at ESO for providing
computer facilities.
This work was supported by the TMR Network ``Gravitational Lensing:
New Constraints on Cosmology and the Distribution of Dark Matter'' of
the EC under contract No. ERBFMRX-CT97-0172, the
``Sonderforschungsbereich 375-95 f\"ur Astro--Teil\-chen\-phy\-sik"
der Deutschen For\-schungs\-ge\-mein\-schaft, and a PROCOPE grant
No. 9723878 by the DAAD and the A.P.A.P.E.
\section*{Appendix: The SkyMaker program}
SkyMaker is an image simulation program, a kind of ``virtual telescope'' originally
designed to assess SExtractor detection and measurement performances (Bertin \& Arnouts
1996). The code (Version 2.3.3\footnote{Freely available at
{\tt ftp://ftp.iap.fr/pub/from\_users/bertin/skymaker/}}) has been much improved since,
and is currently capable of simulating star and galaxy images with higher accuracy.

\subsection*{The optical Point Spread Function (PSF)}
Simulated images feature a PSF which is assumed to be the convolution of the telescopic
(instrumental) point-spread function, tracking errors, and atmospheric ``seeing''
(assuming a long exposure time). We shall however more conveniently deal with the Optical
Transfer Function (OTF), which is just the Fourier transform of the PSF.

\subsubsection*{Telescope}
Although the diffraction spot of a large telescope is small compared to the atmospheric
seeing disk at optical wavelengths, features such as diffraction spikes (created by the
spider-arms of the secondary mirror/primary focus) are obvious on real images and must
be taken into account.

\begin{figure*}[ht]
  \centerline{\psfig{figure=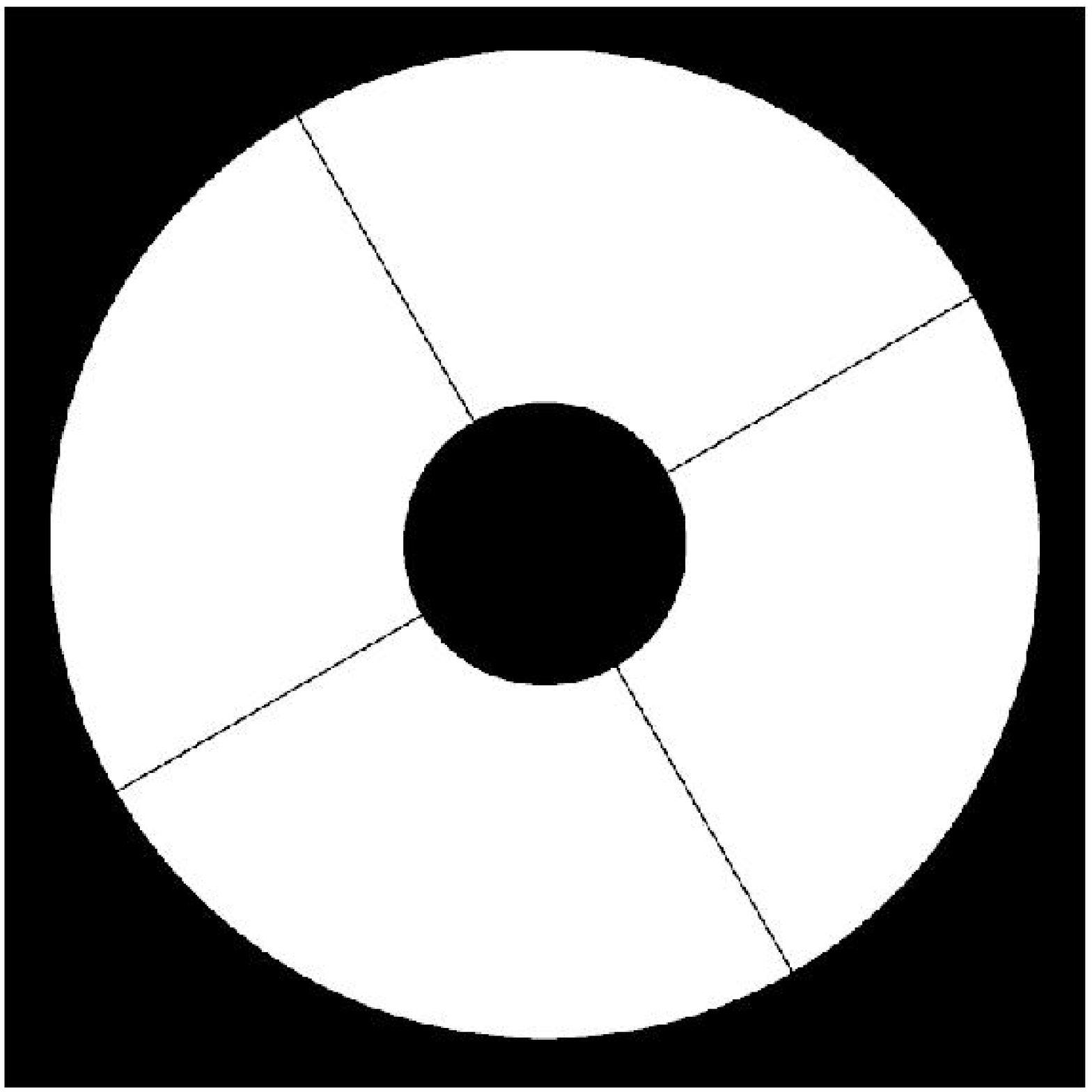,width=0.5\textwidth}
              \psfig{figure=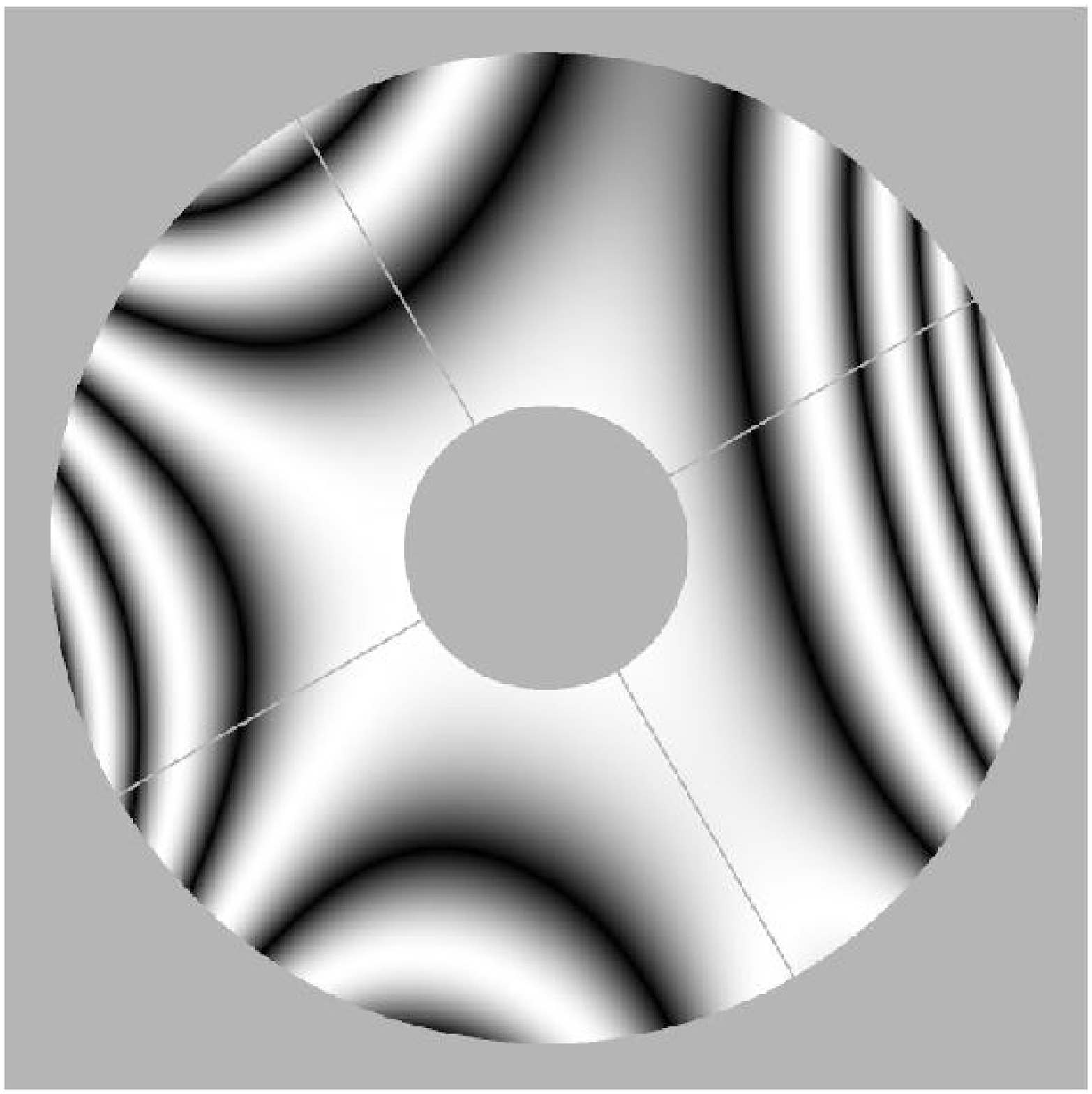,width=0.5\textwidth}}
  \caption{The input pupil used for the simulations. Note the
           secondary-mirror/primary-focus obstruction and the position angle of its support
           arbitrarily set to 30 deg. with respect to image axes.
           {\it Left:} Pupil function modulus.
           {\it Right:} Real part of the pupil function in presence of defocusing, coma
           and triangular aberrations (PSF\_6 from Fig. \ref{psfsfig}).}
  \label{fig:pupil}
\end{figure*}

The principle of the PSF generation is similar to that of other telescope image simulators
like TIM (Hasan \& Burrows 1995). The system is assumed to be illuminated with
incoherent, quasi-monochromatic planar waves. In those conditions, the instrumental OTF is
proportional to the autocorrelation of the input pupil function of the system
(e.g. Born \& Wolf 1999). The on-axis pupil function $P(\rho, \theta)$ used in this work
is shown in Fig. \ref{fig:pupil}. It is typical of telescopes like the 3.6m CFHT,
harbouring wide-field instruments used for weak-lensing experiments.

Instead of using ray-tracing techniques to simulate off-axis aberrations, we preferred
to approximate the latter by modulating the phase $\phi$ of the complex on-axis input
pupil (as shown in Fig. \ref{fig:pupil}). Aberrations are limited to the sum of low-order
Seidel (1856) terms: defocus $\phi_d \propto \rho^2$; astigmatism $\phi_a \propto \rho^2
\cos^2 (\theta-\theta_a)$; coma $\phi_c \propto \rho^3 \cos (\theta-\theta_c)$;
spherical $\phi_s \propto \rho^4$; plus the triangular and quadratic aberrations. Each
term is normalised in ``{\tt d80}'' units, that is, the diameter of the disk within
which 80\% of the light from a point source image is enclosed, omitting the
contribution from diffraction and other aberration terms (Fouqu\'e and Moliton 1996,
private communication).

\subsubsection*{Tracking errors}
A tracking drift of angle $d$ in the $\theta_T$ direction can be simulated by
multiplying the OTF by ${\rm sinc}\,(df).\delta(\theta - \theta_T)$, where $f$ is the
angular frequency. ``Jittering'' effects are easily
added by multiplying the OTF by a centered, 2D Gaussian window.

\subsubsection*{Atmospheric turbulence}
Close to zenith, atmospheric turbulence blurs long-exposure images in an isotropic
way. Experimentally, the ``atmospheric'' OTF is found to quite accurately follow the
theoretical prediction $\propto \exp -3.442\left(\frac{\lambda f}{r_0} \right)^{-5/3}$ for
a turbulent atmosphere that follows Kolmogorov statistics (see for instance Roddier 1981
and references therein, or more recently Racine 1996). Fried's (1966) $r_0$ parameter can
be more conveniently expressed as a function of the ``atmospheric'' Full-Width at
Half-Maximum and the wavelength $\lambda$: $r_0 \approx 0.976 \frac{\lambda}{\rm FWHM}$.
There is no simple analytical expression for the resulting PSF, and although the core
bears some resemblance to a Gaussian, the wings are significantly larger.
The atmospheric FWHM varies only slowly with wavelength ($r_0 \propto \lambda^{6/5}$, i.e.
${\rm FWHM} \propto \lambda^{-1/5}$). This results in a negligible dependency of
star profiles on their spectral characteristics under ground-based standard broadband
observing conditions: something essential for weak-lensing calibration, as stressed by
Kaiser (1999).

\subsubsection*{The Aureole}
A so-called ``aureole'' is observed to dominate all optical instrumental PSFs out to
distances a few times the FWHM away from the centre. This is a feature which must be
taken into account when simulating deep and wide galaxy fields, as it reproduces the
background variations found on real images around bright stars. Although several sources
contribute to the presence of the PSF aureole (light scattering caused by
aerosols, dust on optics, scratches and micro-ripples on optical surfaces, see for
instance Beckers 1995),  it is experimentally found to follow quite a ``universal''
$r^{-2}$ profile (King, 1971). The intensity appears fairly constant too, being close to
16th mag per sq.arcsec for a 0th magnitude star. We adopted this value here.

\subsubsection*{Pixel footprint}
Finally, the generously sampled (${\rm FWHM} \approx 40$ samples) instrumental+atmospheric
PSF is convolved with the square pixel footprint (0.206'', like the CFH12K camera) and
ready to be interpolated over the final image grid.

\subsection*{Simulated stars}
Stars are simulated within SkyMaker, assuming a constant slope of the logarithm of
differential number counts of 0.3 per magnitude interval, and a total sky density of
42000 per sq.degree down to $I$=25. The fairly high slope provides a crude match to star
counts around $I$=20-21 at high galactic latitude (e.g. Nonino et al. 1999), while reducing the density of brighter
stars.

\subsection*{Simulated galaxies}
SkyMaker galaxies are made of a bulge with an $\exp r^{-1/4}$ profile and an exponential
disk. All galaxy parameters are provided by the Stuff program
as described in Sect. \ref{stuffsec}. Obviously,
a study of weak-shear systematics requires accurate profiles, which means
that simulated galaxies must be well sampled. Except for the most extended ones, the
sampling step is that of the PSF (a few mas). Prior to convolution by the PSF,
appropriate thresholding is applied to the wings of the composite galaxy profile to
remove any ``boxy'' limit that may affect shear measurements.

\subsection*{Final image}
A uniform sky background with surface brightness $\mu_I = 19.9$~mag.arcsec$^{-2}$ is
added to the image. Surface brightnesses are then converted to ADUs assuming the image
is the average of twelve 600s CFH12K exposures, yielding a magnitude zero-point of 32.6
mag and an equivalent conversion factor of 36 e$^-$/ADU. Poissonian photon white noise
and Gaussian read-out white noise realisations are eventually applied. It is important
to note that the simulation procedure skips three important steps, present in the
reduction of real images: field warping, image resampling and co-addition. These points
can have a significant impact on weak-shear measurements and will be addressed in a future
paper.

\end{document}